\documentclass[lettersize,journal]{IEEEtran}

\usepackage{cite}
\usepackage{amsmath,amssymb,amsfonts}
\usepackage{algorithmic}
\usepackage{graphicx}
\usepackage{textcomp} 
\usepackage{xcolor}
\usepackage[table]{xcolor}
\usepackage{soul}
\usepackage[export]{adjustbox}
\usepackage{enumitem}
\usepackage{multirow}
\usepackage{float}
\usepackage{pifont}
\usepackage[caption=false]{subfig}

\makeatletter
\renewcommand{\p@subfigure}{\thefigure--}
\makeatother
\usepackage[binary-units]{siunitx}
\def\BibTeX{{\rm B\kern-.05em{\sc i\kern-.025em b}\kern-.08em
    T\kern-.1667em\lower.7ex\hbox{E}\kern-.125emX}}
\usepackage{caption}
\usepackage{subcaption}
\usepackage{threeparttable}
\usepackage{xcolor}      
\usepackage{colortbl}    
\usepackage{array}       
\usepackage{booktabs}
\usepackage{graphicx}
\usepackage{bbding}
\usepackage{pifont}
\usepackage[ruled,vlined,linesnumbered]{algorithm2e}
\usepackage{setspace} 
\usepackage{threeparttable}
\newcommand{\ms}[1]{{\color{black}#1}}
\newcommand{\mv}[1]{{\color{black}#1}}

\begin{document}

\title{SparseCol: A 1320 BTOPS/W Precision-scalable NPU Exploiting Training-free Structured Bit-level Sparsity and Dynamic Dataflow}

\author{
\IEEEauthorblockN{
}
}

\author{
\IEEEauthorblockN{Man Shi, Vikram Jain, Weijie Jiang, Chao Fang, Antony Joseph, Wim Dehaene, Marian Verhelst}
\thanks{Man Shi, Vikram Jain, Weijie Jiang, Chao Fang,  Wim Dehaene, Marian Verhelst are with the ESAT, KU Leuven, Leuven, Belgium (corresponding author: Man Shi, Email: manshi@berkeley.edu).}
\thanks{
Antony Joseph is with NXP Semiconductor, Leuven, Belgium.}
}

\maketitle

\begin{abstract}
Bit-serial computation enables sequential processing of data at the bit level, providing several advantages, such as scalable computational precision. This approach has gained significant attention, especially for exploiting bit-level sparsity in AI workloads. While current bit-serial processors leverage bit-level sparsity to eliminate the computation associated with zero bits, they face a fundamental trade-off: either they suffer from low \ms{memory-access} and computation efficiency caused by irregular patterns of non-zero bits, or they incur substantial area overhead from complex online scheduling mechanisms required to reorganize bit-level data and preserve memory access and computation regularity.

Therefore, we present the \textbf{SparseCol} processor, designed to harness extensive bit sparsity while maintaining high hardware utilization across various AI applications, including CNNs, RNNs, and transformers. In contrast to traditional methods, \textbf{SparseCol} exploits structured bit-level sparsity, \mv{denoted by} bit-column sparsity, without requiring \mv{any re-}training. 
Furthermore, SparseCol implements a dynamic dataflow architecture that tackles hardware under-utilization issues commonly found in existing bit-serial solutions. 
Fabricated in 16nm \mv{CMOS} node, SparseCol delivers 1320 BTOPS/W (BTOPS represents Binary Tera-Operations Per Second, calculated as $\#Wbits \times \#Abits \times TOPS$) peak efficiency while maintaining accuracy, outperforming SotA sparse processors \mv{in terms of efficiency by} $6.8\times$. 
Comprehensive evaluations on CNN classification tasks and transformer architectures demonstrate system-level efficiencies of 745.02 BTOPS/W and 850.5 BTOPS/W, respectively.


\end{abstract}


\section{Introduction}
Deep learning algorithms have achieved state-of-the-art (SoTA) accuracy across numerous AI applications and have become essential tools for addressing complex real-world challenges. However, mainstream predictive neural networks, including convolutional neural networks (CNNs), and generative AI models such as transformers, are both characterized by continuously expanding model sizes and architectural complexity. This growth substantially impedes \ms{the} efficient deployment on computing devices, particularly edge platforms that operate under stringent resource, power, and latency limitations.

\begin{figure}[t]
    \centering
    \includegraphics[width=0.85\columnwidth]
    {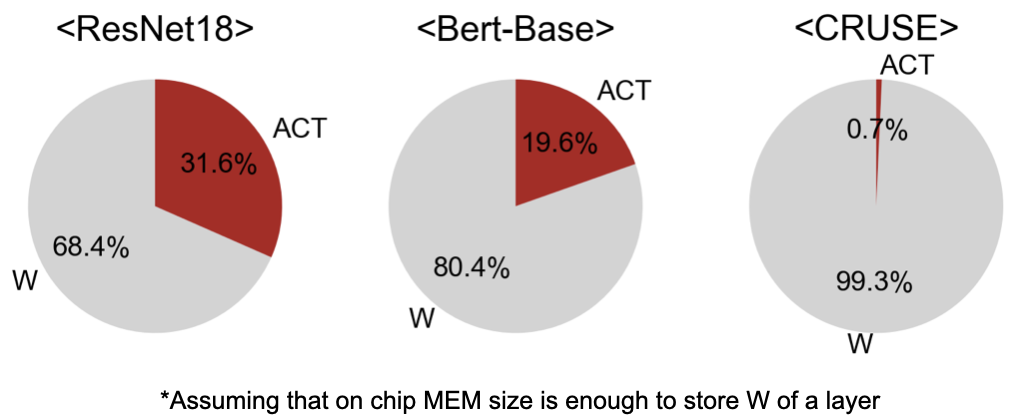}
    \caption{External memory access breakdown for ResNet18, Bert-Base, and CRUSE inference with batch size 1.}
    \label{fig:fig1}
    \vspace{-0.4cm}
\end{figure}

Specialized AI accelerators \cite{jouppi2017datacenter, lu2021sanger, ueyoshi2022diana, tong2024feather, yuan2024hyctor}  have been developed to improve the performance and computational efficiency of neural network inference. Nevertheless, data movement operations emerge as the primary source of system power consumption in AI hardware, with external memory accesses being particularly costly~\cite{9218522}. Fig.~\ref{fig:fig1} illustrates the breakdown of external memory access patterns across different workloads: ResNet18 (CNN)~\cite{he2016deep}, Bert-Base (Transformer)~\cite{vaswani2017attention}, and CRUSE (CNN-LSTM hybrid architecture)~\cite{russakovsky2015imagenet}. Weight-related memory usage dominates all evaluated cases, with \ms{CRUSE} exhibiting the most extreme scenario where weight-related external memory accesses constitute approximately 99\% of total memory traffic.
Consequently, reducing weight storage requirements becomes critical for minimizing both computational overhead and associated memory access demands during inference deployment, which forms the primary focus of this paper. 
 
Weight quantization stands as one of the leading approaches for reducing model size and computational demands. This technique typically falls into two main categories: \mv{In post-training quantization (PTQ)~\cite{liu2021post}, the precision of a model's weights is reduced after a full precision training procedure, while in} quantization-aware training (QAT)~\cite{nagel2022overcoming}, \mv{models are trained at the reduced precision itself}. QAT has demonstrated the capability to aggressively compress parameters to low precision, including 2-bit and 1-bit configurations, while preserving model accuracy.
Yet, QAT presents \ms{several} significant limitations~\cite{choi2016towards}: First, it demands substantial GPU computational resources for model retraining or fine-tuning processes. Second, it raises critical privacy concerns since QAT typically requires access to original training datasets or changing the training method. In contrast, PTQ eliminates the retraining requirement, \mv{yet, at the cost of} 
a more severe trade-off between model compression ratio and accuracy~\cite{liu2021post, xiao2023smoothquant, yao2022zeroquant, wei2023qdrop}. Research findings~\cite{keller202395} indicate that PTQ achieves relatively modest compression gains; for instance, studies have shown that the Bert-Base model can reach approximately $2\times$ compression while experiencing around 1.5\% accuracy loss.

\begin{figure}[t]
    \centering
    \includegraphics[width=0.85\columnwidth]
    {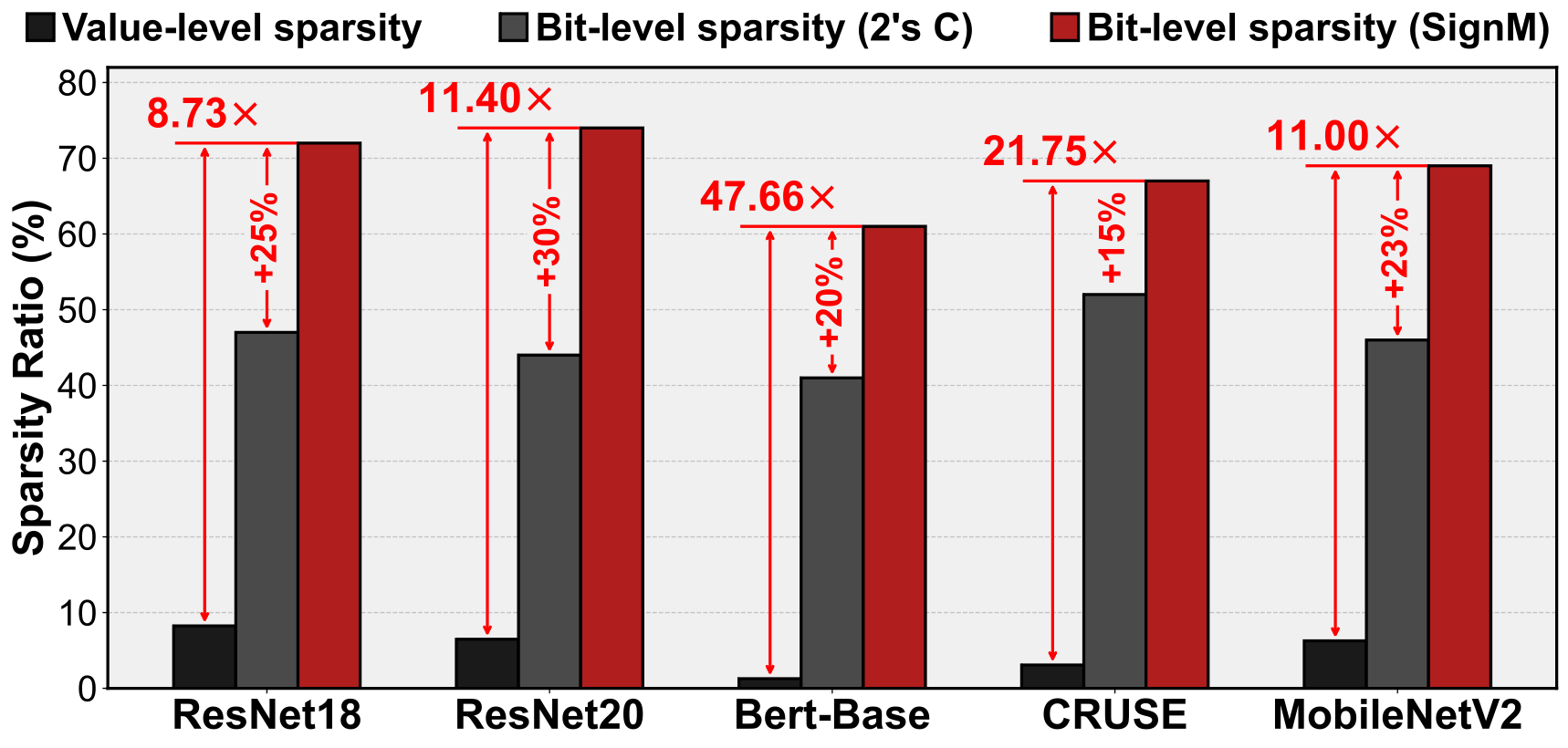}
    \caption{The sparsity ratio of weight at value-level and bit-level, where the binary representations are under 2's complement (2's C) and Sign Magnitude (SignM), respectively.}
    \label{fig:fig1_1}
    \vspace{-0.4cm}
\end{figure}

To further enhance the efficiency, sparsity is widely exploited in AI accelerators~\cite{bitlet, li2022ristretto, gondimalla2019sparten, parashar2017scnn} to skip ineffective operations associated with zero operands. Unfortunately, the advent of new, leaner neural networks and transformer topologies \mv{drastically reduces the model's inherent weight value sparsity.} 
As shown in Fig~\ref{fig:fig1_1}, \mv{well-known} quantized PyTorch models using int8 precision typically show less than 10\% \mv{weight} value-level sparsity without any sparsity-induced training methods. \mv{This is in} sharp contrast \mv{with} bit-level sparsity (BLS) in standard 2's complement binary representation, \mv{which} shows \mv{a sparsity which is} one order larger. 
Furthermore, our findings indicate that adopting a sign-magnitude (SignM) representation yields even more BLS, bringing higher potential performance and efficiency gains.   



Therefore, recent studies~\cite{bitlet,
delmas2019bit, sharify2019laconic, yang2021fusekna, 10946730} have investigated BLS for efficient neural processing. Yet, there are significant challenges to leveraging such abundant BLS effectively from both computational and memory perspectives. 
While initial BLS-aware sparse processors~\cite{delmas2019bit} \ms{demonstrated} promising power and throughput improvements due to bypassing the unnecessary computations whenever a specific (weight) bit is zero, 
they fail to exploit the BLS fully, due to the \textit{irregularity \mv{of memory access and compute}}. 
\mv{As a result,} many recent works \mv{have} proposed \mv{methods} to overcome the irregularity of BLS: BitCluster \cite{li2022bitcluster}, BitPruner \cite{zhao2020bitpruner}, and Bit-balance \cite{sun2023bit} adopt hardware-software co-design to induce structured BLS \mv{during model} training. This, \mv{however,} again encounters the same limitations as QAT. Alternatively, runtime-based methods \cite{delmas2019bit, yang2021fusekna, bitlet} implement complex online hardware bit-schedulers to dynamically balance workloads and maximize BLS utilization,
coming with significant area and power overhead.  
An additional bottleneck \textit{hardware underutilization} further constrains BLS performance: 
BLS-based processors commonly employ bit-serial MAC operations which require more clock cycles per MAC, necessitating proportionally more PEs to match the throughput with bit-parallel MAC. Yet, scaling up PE arrays typically suffers
more severe hardware underutilization~\cite{10119180}. 

To address these limitations, we introduce bit-column sparsity, a novel approach that exploits structured BLS in SignM binaries without requiring specialized training methods. This approach \mv{can be further enhanced with} a post-processing 
strategy (Bit-Flip) that boosts bit-column sparsity in pre-trained models within 1\% drop. We \mv{finally exploit} this concept \mv{in} \textit{SparseCol}, a structured BLS-aware neural processor that synergistically integrates bit-column-based computation and compression with configurable dataflow to achieve high throughput and energy efficiency. 
The contributions of this work are hence as follows:

\begin{itemize}
	\item Leverage bit-column sparsity for efficient computation and compression without training involvement.
    \item Explore the dynamic dataflow support with the tradeoff between flexibility and area overhead. 
    \item Implement a precision-scalable and efficient DNN processor, \textit{SparseCol}, with a compatible architecture to efficiently leverage bit-column sparsity and dynamic dataflow processing.
\end{itemize}


Sec.~\ref{sec:background} introduces the \mv{hardware-algorithm co-design fundamentals on which this work is built, encompassing the relevant} dataflow concepts \mv{and} 
the proposed bit-column sparsity approach. Sec.\ref{sec:core1} and \ref{sec:core2} first present the detailed design of the two main computing engines, followed by the \textit{SparseCol} overall system architecture  (Sec.~\ref{sec:system_architecture}) elaboration.
Sec.~\ref{sec:measurements} finally presents a comprehensive evaluation \mv{of the \ms{taped-out} chip}, demonstrating a peak energy efficiency of 1320 BTOPS/W with a $6.8\times$ efficiency \ms{improvement} over existing solutions.

\begin{figure*}[t]
    \centering
    \includegraphics[width=1.95\columnwidth]
    {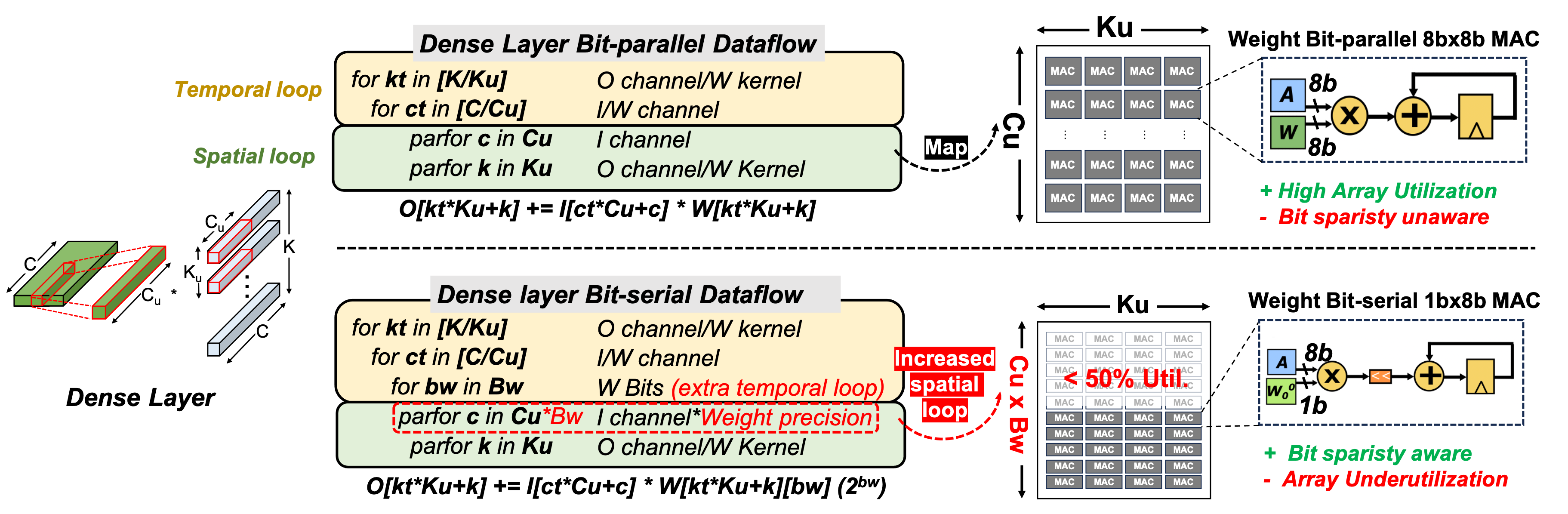}
    \caption{Top: Weight bit-parallel dataflow, MAC array and MAC;
 Bottom: Weight bit-serial dataflow, MAC array and MAC.}
    \label{fig:fig2}
    \vspace{-0.4cm}
\end{figure*}

\section{Background and motivation}
\label{sec:background}

This section explores the algorithm-hardware co-design principles \mv{of \textit{SparseCol}}, \mv{starting with the necessary background on dataflow, followed by the proposed bit-column} 
sparsity approach.
\begin{figure}[t]
    \centering
    \includegraphics[width=0.8\columnwidth]
    {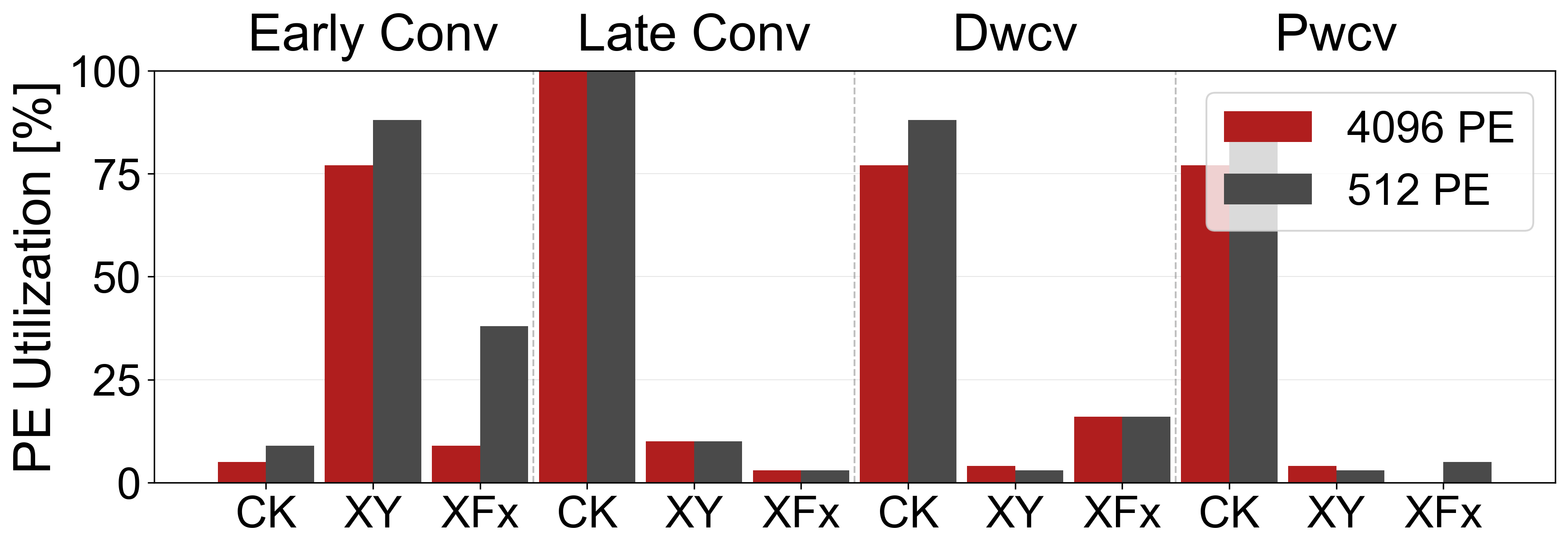}
    \caption{\ms{Layer-wise PE utilization evaluation with different SUs.}}
    \label{fig:fig4c}
    \vspace{-0.4cm}
\end{figure}
\subsection{Dataflow concepts}
Neural network layers \mv{can be represented with} nested \mv{for-}loop structures. For instance, a CNN layer computation typically requires iterating through multiple dimensions: the output feature map parameters (OX, OY, K, B), input feature map dimensions (IX, IY, C), and weight kernel configurations (FX, FY, C). 
Hardware efficiency is achieved by strategically reordering and partitioning loops to create different dataflow implementations~\cite{10970753}. 
\ms{A careful balance of spatial unrolling (SU) and temporal unrolling (TU) is important in this optimization}. Spatial unrolling enables concurrent processing by parallelizing loop dimensions within a single clock cycle. \mv{Practically, this} determines how inner loops are \mv{spatially} distributed across the PE array for simultaneous computation. 
Temporal unrolling processes the remaining loop dimensions sequentially. 
The combination of the temporal tiling of the nested loops and their spatial unrolling provides the dataflow of a given workload~\cite{10129330}.

Fig.~\ref{fig:fig2} exemplifies possible execution dataflows of a dense layer under two computational manners: weight bit-parallel~\cite{genc2021gemmini} and bit-serial~\cite{albericio2017bit} dataflows. In bit-parallel computation, all bits of the weight are processed simultaneously in parallel. As shown in Fig.~\ref{fig:fig2}(top), the workload is partitioned 
over tiles of channels ($C_u$ and $K_u$, \mv{in which the suffix 'u', indicates the spatial unrolling dimensions}). \mv{This allows them to be efficiently} mapped onto a MAC array \mv{of dimension $C_u*K_u$} where each MAC operates on full-precision (for example, 8-bit) inputs and weights, resulting in high array utilization. However, because each MAC processes all bits of the weights regardless of their binary value, this approach is unaware of BLS, meaning it performs unnecessary operations on zero bits.
In bit-serial computation (Fig.~\ref{fig:fig2}(bottom)), weights are processed one bit at a time, serially across time steps. This introduces an additional temporal loop over weight bits (e.g., $b_w$ in $B_w$), and expands the spatial loop size by a factor of the weight bitwidth ($C_u * B_W$) to \mv{achieve} iso-throughput compared to bit-parallel computation. This \mv{temporal unrolling of} precision allows the architecture to leverage BLS through zero-bit skipping across clock cycles. \ms{However, finer-granularity management requires larger PE arrays, making it difficult to efficiently map workloads with varying tensor sizes. This results in substantial PE idling and degraded utilization. Prior work \cite{du202328nm} shows that optimal dataflow mappings depend on layer geometry: XY-parallel SUs favor wide layers, CK-parallelism suits deep layers, and XFx-parallelism benefits large-kernel layers. Specifically, we analyze PE utilization across four representative workload types from MobileNetV2 and ResNet18: early layers (wide, shallow), late layers (narrow, deep), depthwise convolutions (single channel), and pointwise convolutions (1×1 kernels), on both 4096 1b×8b and 512 8b×8b PE arrays under each mapping configuration as shown in Fig.~\ref{fig:fig4c}. 
This analysis reveals a critical bottleneck: static PE arrays cannot maintain high utilization across diverse layer shapes, especially for larger array size. While current bit-serial solutions typically overlook this limitation, our work addresses it through dynamic dataflow adaptation, as detailed in subsequent sections.} 



\begin{figure}[t]
    \centering
\includegraphics[width=1\columnwidth]
    {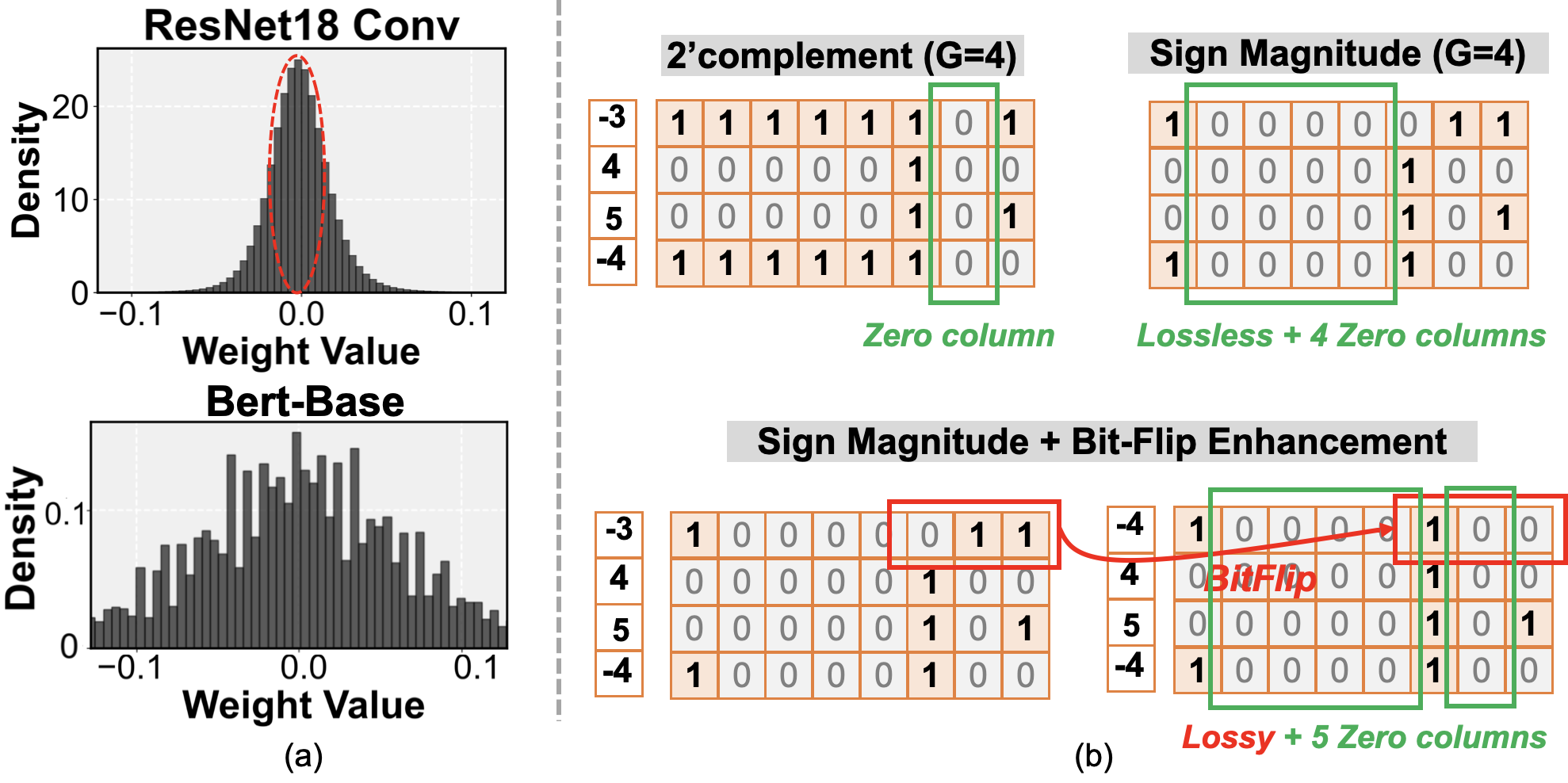}
    \caption{(a). ResNet18 quantized weights under normal distribution and Bert-Base quantized weights under uniform distribution. (b) Bit column sparsity under 2's complement and sign magnitude representations, as well as bit-flip enhancement. }
    \vspace{-0.4cm}
    \label{fig:fig3}
\end{figure}

\subsection{Bit-column sparsity}
To better leverage the BLS, we examine the distribution and binary representation of weights for different models. 

On the one hand, many quantized neural networks, like ResNet18, exhibit weight distributions heavily concentrated around zero, as shown by the normal distribution in Fig.~\ref{fig:fig3}(a) top. While this intuitively suggests opportunities for sparsity exploitation, the commonly used 2's complement binary creates highly irregular BLS patterns that are difficult to leverage efficiently. To address this irregularity, we introduce two modifications: 1.) \ms{operate in the sign-magnitude domain \cite{10067269}, increasing both the amount and locality of sparsity without any accuracy degradation, as weights are typically symmetrically quantized \cite{wu2020integerquantizationdeeplearning, xiao2023smoothquant}}; and 2.) introduce the concept of bit-column sparsity. \ms{For bit-column sparsity, weight values are grouped with a configurable group size G. Within each group, we analyze the same bit position across all weights to identify columns where all bits are zero, creating exploitable vertical sparsity patterns.} 
When an entire column within the weight group contains only zero bits, we define this as a "zero column," representing exploitable bit-column sparsity. As demonstrated in \mv{Fig.~\ref{fig:fig3}(b) left,} in a 2's complement notation, the number of naturally occurring zero columns is extremely limited. 
However, as symmetric quantization~\cite{kulkarni2022survey} is widely adopted in weight quantization schemes, we can losslessly convert weight format from 2's complement to sign-magnitude. \mv{As seen in Fig.~\ref{fig:fig3}(b) right,} this transformation significantly improves bit-column sparsity characteristics. 
\begin{figure*}[t]
    \centering
\includegraphics[width=1.8\columnwidth]{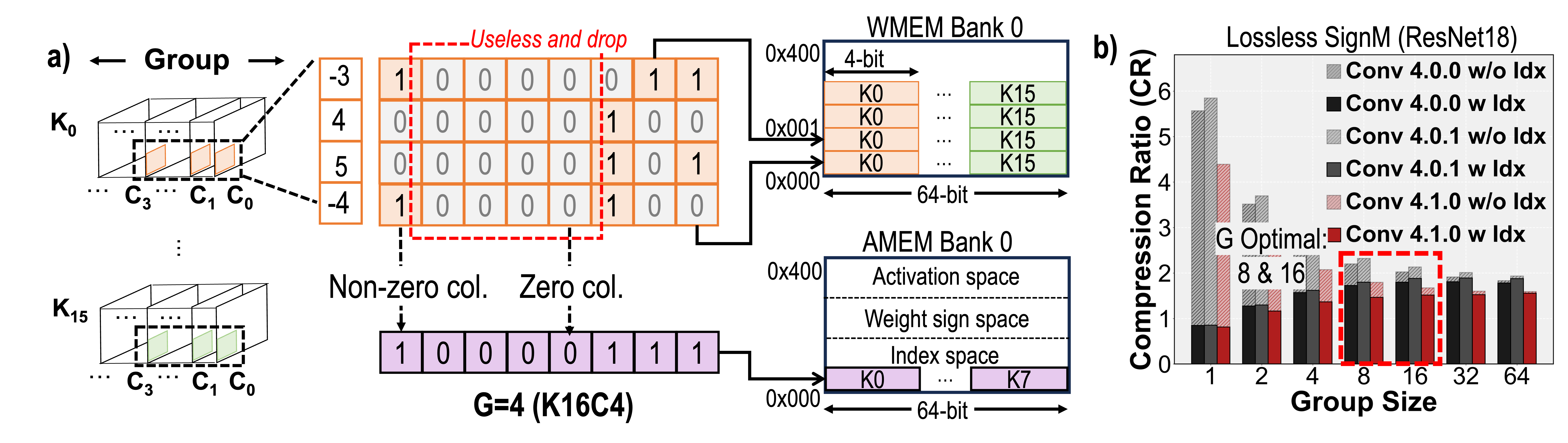}
    \caption{(a) Sparse-column based compression with adaptive activation memory space partition. (b) The tradeoff between group size and compression ratio for ResNet18, \ms{where conv 4.0.0 w/o Idx" stands for "conv 4.0.0 without index overhead" and "conv 4.0.0 w/ Idx" stands for "conv 4.0.0 with index overhead included.}
    \ms{All groups within the layer are uniformly padded to four non-zero bit columns to enable regular memory access in this case.}}
    \label{fig:fig6}
    \vspace{-0.4cm}
\end{figure*}
On the other hand, in some networks, such as Bert-Base, with near-uniform weight distributions, even sign-magnitude representation yields limited bit-column sparsity. \mv{Here}, we introduce the Bit-Flip enhancement strategy. This approach strategically flips certain bits to create additional zero columns while maintaining acceptable accuracy. 
In the bottom example of Fig.~\ref{fig:fig3}(b), one more zero column can be provided by flipping the original $011$ to $100$. This Bit-Flip represents a lossy, yet highly effective, enhancement that maximizes hardware efficiency opportunities while preserving model performance within acceptable bounds. 
\ms{While Bit-Transformer~\cite{9570718} also exploits bit-level sparsity through bit-flipping for RRAM-based analog computing, SparseCol is specialized for digital accelerator and achieves higher bit-column sparsity by employing SignM format and enables regular memory access patterns through the layer-wise Bit-Flip strategy.}
\ms{More details on Bit-Flip algorithm optimization can be found in~\cite{shi2024bitwave}, which focuses on algorithm optimization and architectural concepts. This paper offers complementary content by presenting comprehensive silicon implementation details and chip characterization, compared to~\cite{shi2024bitwave, 11074933}. We bridge the gap between theoretical benefits and practical implementation, demonstrating the real-world viability of the proposed techniques.}

\begin{figure*}[tb]
\centering
\includegraphics[width=1.6\columnwidth]{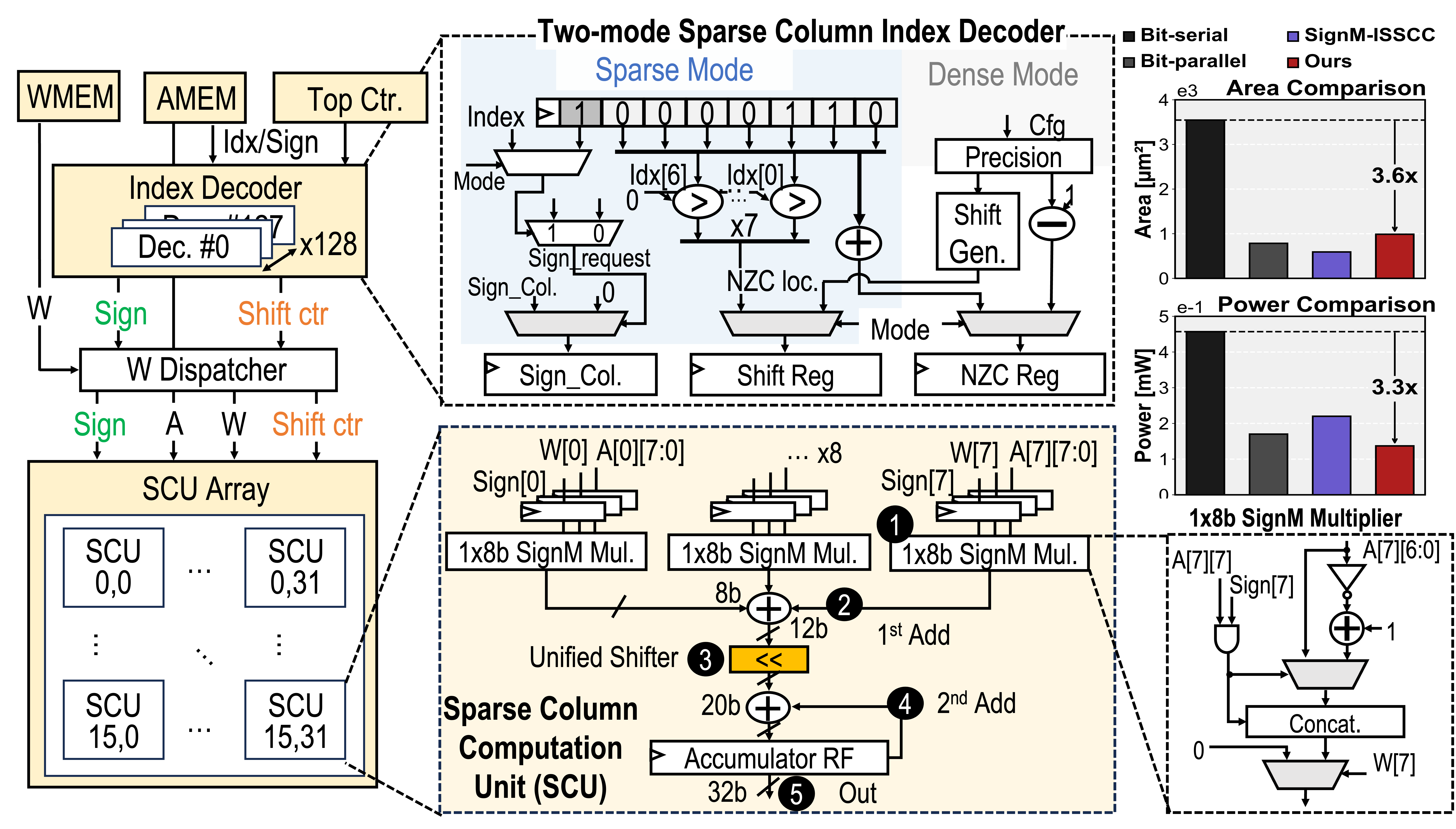}
\caption{\ms{Bit-serial sparse column computation flow and its core component architectures: SCID and SCU. Right Top: Simulated power, area comparison of four PE types: traditional bit-serial, bit-parallel, SignM-isscc\ms{\cite{10067269}}, and \ms{ours} under iso-throughput.}}
\label{fig:bscc}  
\vspace{-0.2cm}
\end{figure*}

\section{Processing engine}
\label{sec:core1}
This section presents a detailed analysis of the processing engine architecture, which is the main component of \textit{SparseCol} that features three key innovations to leverage the proposed bit-column sparsity: (1) Sparse-column-based weight compression; (2) Bit-serial sparse column computation; (3) Dynamic dataflow processing.

\subsection{Sparse-column-based weight compression}
The first feature of the processing engine is \mv{its ability to directly operate on a compressed weight representation, building on top of the introduced bit-column sparsity.} 
As shown in~\ref{fig:fig6}(a), \mv{the sparse-column-based weight compression} is realized by grouping weights along the input channels ($C$) in groups of $G$ weights, and identifying columns of same significance zero-bits in each group. \mv{An 8-bit group index vector, indicating for each group the location of zero and} Non-zero columns (NZCs). \mv{Subsequently, only the NZC are} 
stored in \mv{the WMEM} memory \mv{using} a bit-first \mv{memory layout}~\cite{kim2016bit} \mv{grouping} multiple output channels per memory row. \ms{To simplify hardware implementation, all weight groups within a layer are uniformly flipped to have the same number of non-zero bit columns, determined by the Bit-Flip strategy. This ensures regular memory access patterns and simple control logic at the cost of a modest reduction in compression ratio.}
In addition, we smartly arrange the weight index and non-zero sign bits in the activation memory (AMEM) to further increase the memory efficiency, which will be discussed in Sec.\ref{sec:mem} in depth.
The group size ($G$) clearly influences the trade-off between the total memory footprint and the memory utilization, as a too large $G$ reduces the compression effectiveness, while a too small $G$ requires more indexing and complicates data layout of each memory row. 
Experimental analysis (Fig.~\ref{fig:fig6}(b)) identifies optimal $G$ values of 8 and 16, while G=32 and G=64 achieve similar overall compression ratios. \ms{This is because smaller group sizes offer two advantages over larger groups despite similar compression ratios. First, they reduce memory access granularity, benefiting layers with smaller kernels by avoiding unnecessary data retrieval. Second, they better align with our spatial unrolling introduced in Section III: since weights from the same group are stored contiguously as one non-zero column, large groups (e.g., G=64) force retrieval of entire 64-element blocks even when only partial data is needed. G=8 and G=16 thus balance compression efficiency with memory access costs.}


\subsection{Bit-serial sparse column computation}
\label{sec:datapath}
The processing engine's {datapath} enables direct computation on the compressed weights from WMEM without requiring decompression, implementing what we term bit-serial sparse column computation (BSCC). Fig.~\ref{fig:bscc} presents BSCC computation flow and the micro-architectures for the core modules: sparse column index decoder and sparse column computation unit. 
\subsubsection{Sparse column index decoder} 
To support BSCC, a sparse column index decoder (SCID) is needed first to parse the index information and ensure computation correctness. 
Fig.~\ref{fig:bscc} depicts \mv{the} 8-bits wide SCID architecture, which operates in two distinct modes: sparse and dense processing. 

During the sparse mode operation, the module automatically generates control signals by \mv{deriving the location of the NZC's from the} index vector. In the example of Fig.~\ref{fig:bscc}(top), the 8-bits index is partitioned into two components: the MSB 
and the remaining bits. A "1" at MSB triggers the requesting access to the NZ sign bit column, otherwise the zero sign column is generated locally. The remaining index bits $Idx[6:0]$ encode the sparsity information for individual data-bit columns.
In practical workload, $Total_{index}$ index bits are fetched per memory access, which depends on three factors: weight precision ($W_p$), weight-related spatial unrolling factors ($K_u$, $C_u$), and group size ($G$). 
Firstly, based on the sparse-column compression, $W_p$-length indexes are demanded by one weight group and hence packed together to ensure parallel process, as shown in Fig.~\ref{fig:bscc}(top). 
Then, $\frac{ K_{u} \times C_{u}}{G}$ presents the number of weight groups that is mapped on the PE array per cycle. The $Total_{index}$ can be formulated as:
\begin{equation}
Total_{index} = W_{p} \times \frac{ K_{u} \times C_{u}}{G}
\end{equation}
Specifically, our system employs INT8 weight compression and supports: $[K_{u}=32, C_{u}=8]$, $[K_{u}=32, C_{u}=32]$, and $[K_{u}=128, C_{u}=8]$ (elaborated in Sec.~\ref{sec:df}), combined with group sizes of 8 or 16. This results in $Total_{index}$ ranging from 128 to 1024 bits. 
To handle this variation efficiently, the complete system incorporates 128 parallel 8-bit SCID units, enabling simultaneous processing of index streams up to 1024 bits wide. Every clock cycle, the SCID extracts the "1"'s position in index sequentially to output $\frac{ K_{u} \times C_{u}}{G}$ unique shifts, which are paired with associated sign lines for later computation.
Notably, the $Total_{index}$ bits can be reused across all NZCs within the same weight group, thereby amortizing the bandwidth demands across multiple clock cycle computations. This property is leveraged to establish the memory access optimization as discussed in Sec.\ref{sec:mem}

For layers with \mv{limited} sparsity, where indexing overhead would exceed the compression benefits, or layers containing dynamically computed operands such as transformer attention mechanisms, the system switches to dense mode. 
In this mode, SCID \mv{operates with} a configurable nominal weight precision ranging from 2 to 8 bits.  
This configuration drives an internal counter that produces control signals without requiring explicit indexing.

\subsubsection{Sparse column computation unit}
Once the shift controls and sign column are prepared by SCID, the actual BSCC takes place in the sparse column computation units (SCU). 

 

SCU exploits the shared-significance property in BSCC by accumulating partial results within each bit-column before applying shift operations. This optimization requires spatial unrolling along input channel (C) or kernel (FX/FY) dimensions across bit-columns. We assume C-dimension unrolling, typically in multiples of eight, to maximize hardware utilization. Rather than shifting individual weight bits, SCU performs a single shift operation per entire bit-column after accumulation, reducing computational overhead. As demonstrated in Fig.~\ref{fig:bscc}, BSCC involves the five-step process within one SCU: \ding{182} each SCU receives eight 8-bit activations, an 8×1-bit weight column, and \mv{8} sign bits and 
one SignM multiplier executes 8×1-bit signM multiplication; \ding{183} partial sum accumulation ($1_{st}$ Add) combines all column element results; \ding{184} \mv{the Shifter} applies the SCID-derived shift information to align the accumulated non-zero column based on its bit significance; \ding{185} the $2^{nd}$ Add conducts the local accumulation and follows the final output generation (\ding{186}). 
Sign bits and activations are reused across multiple cycles until all non-zero columns for the current weight group are processed, while weight bits are updated each cycle.

Fig.~\ref{fig:bscc}(RightTop) shows that the SCU's unified shifter enables efficient group-level shift-then-add computation, offering $3.6\times$ area and $3.3\times$ power reduction compared to the traditional bit-serial MAC at iso-throughput.

\begin{figure*}[tb]
\centering
\includegraphics[width=1.8\columnwidth]{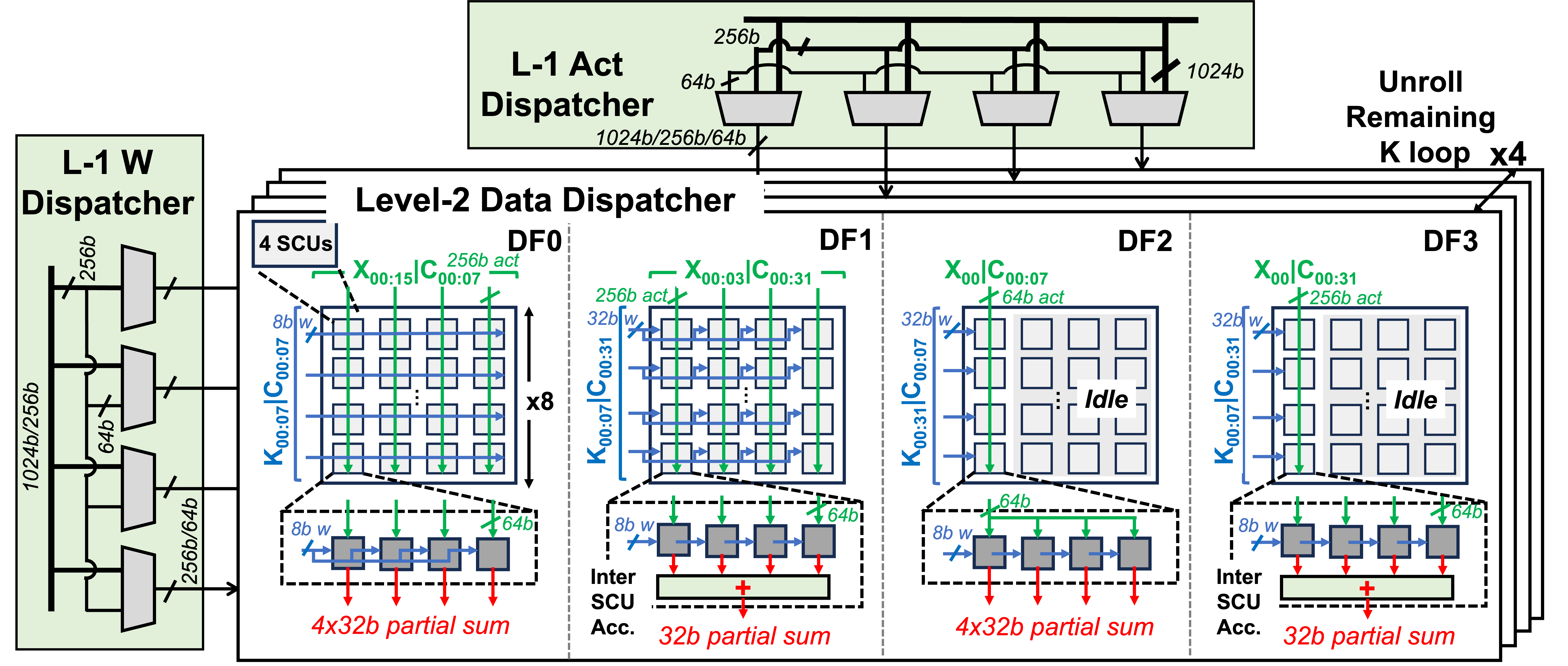}
\caption{\ms{SUC array with optimized four dynamic dataflows, running one dataflow each time}} 
\label{fig:su_4}  
\vspace{-0.3cm}
\end{figure*}

\begin{table}[!t]
    \caption{7 SUs, 4SUs (highlighted in yellow), 1SU (highlighted in red text) and corresponding bandwidth}
    \centering
     \scriptsize
    \begin{threeparttable}
    \resizebox{0.48\textwidth}{!}{
	\begin{tabular}{|c|c|c|c|c|}
		\hline
        ~ & \multirow{ 2}{*}{SUs} & W BW & Act BW   \\
		~ & ~ &  (bit/cc)  & (bit/cc)  \\
		\cline{1-4}
  
		\rowcolor{yellow!75} $SU_0 (DF_0)$ &$[C_u=8, OX_u=16, K_u=32] $& 256 & 1024 \\
		\cline{1-4}

		$SU_1$ &$[C_u=16, OX_u=8, K_u=32] $& 512 & 1024 \\
		\cline{1-4}
           
            \rowcolor{yellow!75} $SU_2 (DF_1)$ & $[C_u=32, OX_u=4, K_u=32]$  & 1024 & 1024 \\
		\cline{1-4}
            \rowcolor{yellow!75} $SU_3 (DF_2)$ & $[C_u=8, OX_u=1, K_u=128]$  & 1024 & 64 \\
		\cline{1-4}
         $SU_4$ & $[C_u=16, OX_u=1, K_u=64]$& 1024 & 128\\
		\cline{1-4}
             \rowcolor{yellow!75} \textcolor{red}{$SU_5 (DF_3)$} & \textcolor{red}{$[C_u=32, OX_u=1, K_u=32]$} &  \textcolor{red}{1024} & \textcolor{red}{256}\\
		\cline{1-4}
		\hline
  $SU_6$ \footnote{1} & $[G_u=64, OX_u=2, K_u=1]$\tnote{1} & 64 & 1024 \\
		\cline{1-4}
		\hline
	\end{tabular}
    }
    \begin{tablenotes}
        \footnotesize
        \item[1] SU specialized for the Depthwise Convolution.
      \end{tablenotes}
    \end{threeparttable}
        \label{table:su}
        \vspace{-0.3cm}
\end{table}

\subsection{Dynamic dataflow processing}
\label{sec:df}
While Section~\ref{sec:datapath} covered a single SCU computation flow, this section addresses the coordination of multiple SCUs to optimize data reuse efficiency, realized by the dynamic dataflow \mv{support} (the third feature of the processing engine).

Firstly, to fulfill the high throughput demands of DNNs, we deploy 512 SCU modules in parallel, encompassing 4096 1b×8b SMMs that deliver a peak performance equivalent to 512 8b×8b bit-parallel PEs. However, this large number of SMMs \mv{bring a risk in terms of} underutilization of \mv{the} PE array across workloads with diverse dimensions. To address this challenge, we equip the PE array with dataflow reconfiguration capabilities \mv{to} dynamically adapt spatial unrolling dimensions based on layer characteristics.
\begin{figure}[tb]
\centering
\includegraphics[width=0.85\linewidth]{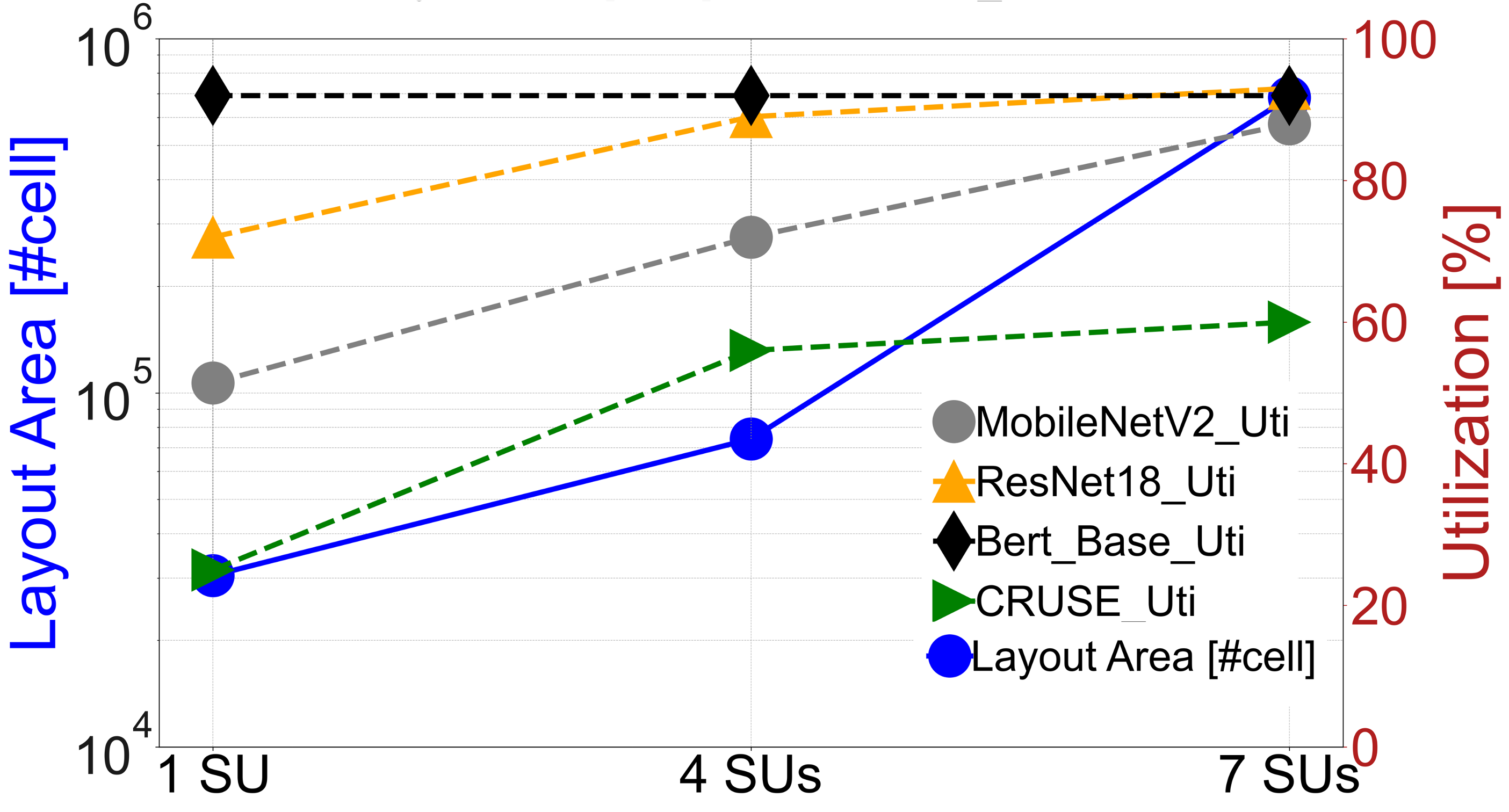}
\caption{PE utilization evaluations for 1SU v.s. 4SUs v.s. 7SUs; Area numbers are obtained from PNR implementation under 0.8V nominal voltage with \ms{238Mhz} frequency. Utilization is actual hardware utilization when mapping specific workloads.} 
\label{fig:su_comp}  
\vspace{-0.4cm}
\end{figure}

To determine the optimal SUs for \textit{SparseCol}, we employed \ms{a} dataflow optimization tool~\cite{mei2021zigzag} for systematic exploration. Various neural network workloads, including CNNs, RNNs, and transformers, were analyzed through the optimizer to evaluate performance and energy consumption across different spatial unrolling configurations. This analysis initially identified seven SU patterns (in Table~\ref{table:su}) that consistently maintain high utilization efficiency across diverse computational workloads, as discussed in~\cite{shi2024bitwave}. Yet, \mv{the hardware support for the combination of all seven spatial unrolling schemes} comes with a significant hardware \mv{overhead}. We \mv{hence} explored the tradeoff between flexibility and hardware cost; four optimized dynamic dataflows were then strategically selected from the original seven SUs. Consequently, 
the \textit{SparseCol} chip implements the dynamic dataflow with the 4 SUs supported (Fig.~\ref{fig:su_4}). 

This exploration is realized through a comparative analysis of area and utilization across three design strategies: 
(1) implementing a single spatial unrolling with minimal hardware cost and flexibility ($SU_5$ in Table~\ref{table:su}), (2) enabling seven spatial unrollings for high utilization across all benchmarks but the most complex NoC (all SUs in Table~\ref{table:su}), and (3) implementing four spatial unrollings that maintains decent utilization but significantly simplify the complexity of flexible data dispatcher (4 SUs, highlighted in yellow in Table~\ref{table:su}).

Fig.~\ref{fig:su_comp} presents layout area (from place and routing (PNR) implementation) and utilization metrics across three design strategies (1 SU, 4 SUs, and 7 SUs) for various neural network architectures. 1 SU implements the single SU configuration that exhibits the highest average hardware utilization across different neural network workloads compared to the remaining six SU configurations. 
Fig.~\ref{fig:su_comp} \ms{demonstrates that the layout area significantly reduces from 7 SUs to single SU, exhibiting approximately a 23-fold reduction. However, this area efficiency comes at a performance cost, as utilization decreases for most workloads. This effect is particularly pronounced for CRUSE\_Uti, where utilization drops from approximately 58\% in 7 SUs to 24.7\% in 1 SU. To identify the optimal configuration, we introduce an area-normalized efficiency metric (hardware utilization / normalized systarea): For CRUSE, 1-SU achieves 0.22, 4-SU reaches 0.24, while 7-SU drops to 0.026. The 4-SU implementation represents the optimal design point. In contrast, scaling to 7-SU incurs $23\times$ area cost for merely 7\% utilization gain. Despite a 17\% utilization drop for MobileNetv2 due to the absence of dedicated depthwise dataflow, the 4-SU configuration achieves the best balance between performance and area cost, requiring less than one-tenth of the area compared to the 7-SU design.}



To better visualize, the total 512 SCUs are arranged in 4 groups of 128. In Fig.~\ref{fig:su_4},
Within a single group, 128 SCUs can be configured to one of four dataflows ($DF_0$ to $DF_3$).  
$DF_0$ and $DF_1$ are engineered for shallow/deep convolution and matrix multiplication operations, while $DF_2$ and $DF_3$ are tailored for narrow/wide dense layers.
Supporting multiple spatial unrollings requires a flexible Network-on-Chip (NoC) to manage different data dispatch patterns among computing units. Therefore, a two-level data dispatcher is designed. 
\mv{As shown in Fig.\ref{fig:su_4}}, Level-1 Dispatchers form the primary distribution network, consisting of specialized modules for weights (L-1 W Dispatcher) and activations (L-1 Act. Dispatcher). 
As summarized in Table~\ref{table:su}, the L-1 W Dispatcher handles weight distribution with \mv{a} 1024 or 256-bit \mv{per cycle bandwidth}, while the L-1 Act Dispatcher manages activation data in 1024/256/64-bit for different SUs across 4 groups of SCUs. 

The Level-2 Data Dispatcher enables fine-grained control within each group (128 SCUs). 
The dispatcher supports various data reuse patterns for activation, weight, and partial sum.
Specifically, examining DF0 as a representative example, 128 SCUs are arranged in an
8 (rows) x 16 (columns, showing in a 4x4 structure) array for better explanation.
Green lines in Fig.~\ref{fig:su_4} illustrate the activation sharing pattern: a single activation $X$ with 64-bit data spanning channels $C_{00:07}$ is broadcast vertically across all rows, while the $X_{00:15}$ parallelism is distributed horizontally across the 16 columns. Blue lines demonstrate the weight sharing pattern: a weight $K$ with 8-bit data for channels $C_{00:07}$ is broadcast horizontally across all columns, while the $K_{00:07}$ parallelism is distributed vertically across the 8 rows.
Examining the 4 SCU subset in detail: each SCU receives 8-bit weights and 64-bit activations, where the same weights are reused with unique activations to generate 4 output features ($OXs$) within a output channel. Consequently, there is no accumulation of intermediate computational results among SCUs within $DF_0$. This contrasts with $DF_1$ and $DF_3$, where outputs from every 4 horizontally adjacent SCUs are accumulated to generate one partial sum, as the $C_{00:31}$ channel parallelism is evenly partitioned across every 4 SCUs. The remaining dataflows can be interpreted through analogous architectural principles.

\begin{figure}[tb]
\centering
\includegraphics[width=0.8\linewidth]{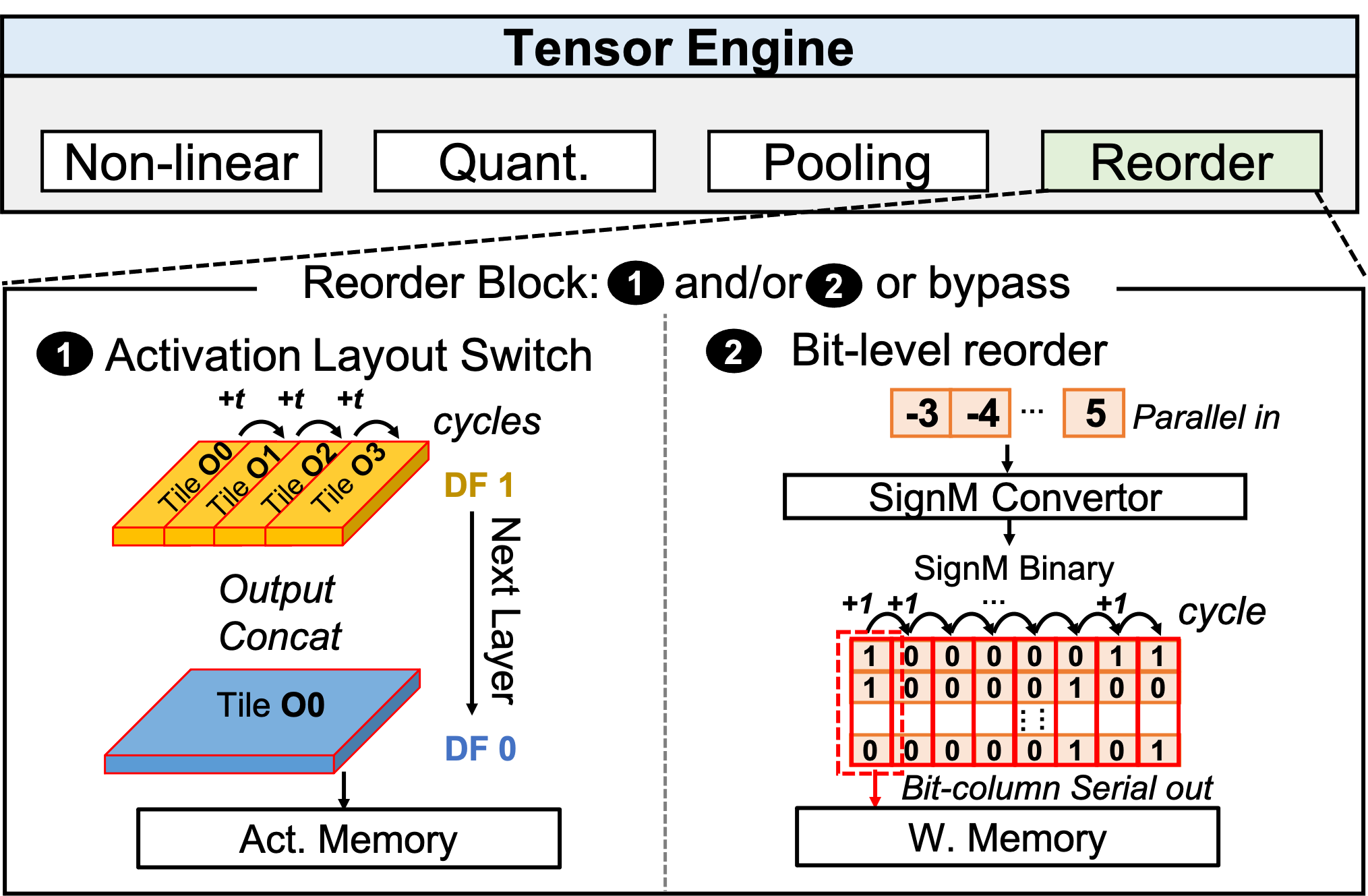}
\caption{Tensor engine with special focus on reorder block.} 
\label{fig:tensor} 
\vspace{-0.2cm}
\end{figure}

\section{Tensor engine}
\label{sec:core2}
The tensor engine handles optional post-processing operations essential for neural network inference. 
As shown in Fig.~\ref{fig:tensor}, the engine integrates four key functional modules. A lookup table-based non-linear module with 64-way segments and 32-bit precision supports various activation functions, such as tanh, sigmoid, and softmax, required by RNNs and transformers. 
An integrated pooling core provides MaxPool and AvgPool operations with instruction-configurable window sizes. Additionally, a quantization block with floating-point computation, featuring 32 FP32 multipliers to enable per-channel and per-tensor scaling for outputs re-quantization.

A particularly noteworthy component is the reorder block, depicted at the bottom of Fig.~\ref{fig:tensor}. 
Specifically, this reorder block with two operational modes is designed to optimize the data layout, aiming to address memory access efficiency and layout compatibility issues.

The first mode is Activation Layout Switch, reformatting output data layout. This mode helps to bridge dataflow mismatches between consecutive layers. 
For example, when a previous dataflow, such as $DF_1$ with small OX parallelism feeds into the next layer's dataflow ($DF_0$), a set of register files in this block buffers the small tiles (Tiles 00, 01, 02, 03) across time until sufficient data concat to match the next layer's parallelism requirements. Once this is met, data is read from these registers in a layout compatible with the subsequent dataflow, enabling efficient single-cycle required data fetching. The reorganized data is then sent to AMEM for next-layer access.


The second mode is Bit-level reorder. It is specifically designed for transformer attention mechanisms. This component first converts bit-parallel data (in 2's complement format) to SignM. Then, the SignM represented data is transformed into a bit-column serial output format and stored in WMEM across multiple clock cycles. This conversion is crucial for transformer architectures as it enables online-generated operands to be reformatted into a bit-column serial data layout compatible with BSCC in subsequent layers. 
To accommodate the data reorder requirements between different SUs and modes, the  Reorder block contains 1KB \ms{register} files. Both modes in this block can be selectively engaged, providing flexibility in the dataflow optimization strategy based on the specific neural network architecture and computation requirements.

\begin{figure}[t]
    \centering
    \includegraphics[width=1\columnwidth]{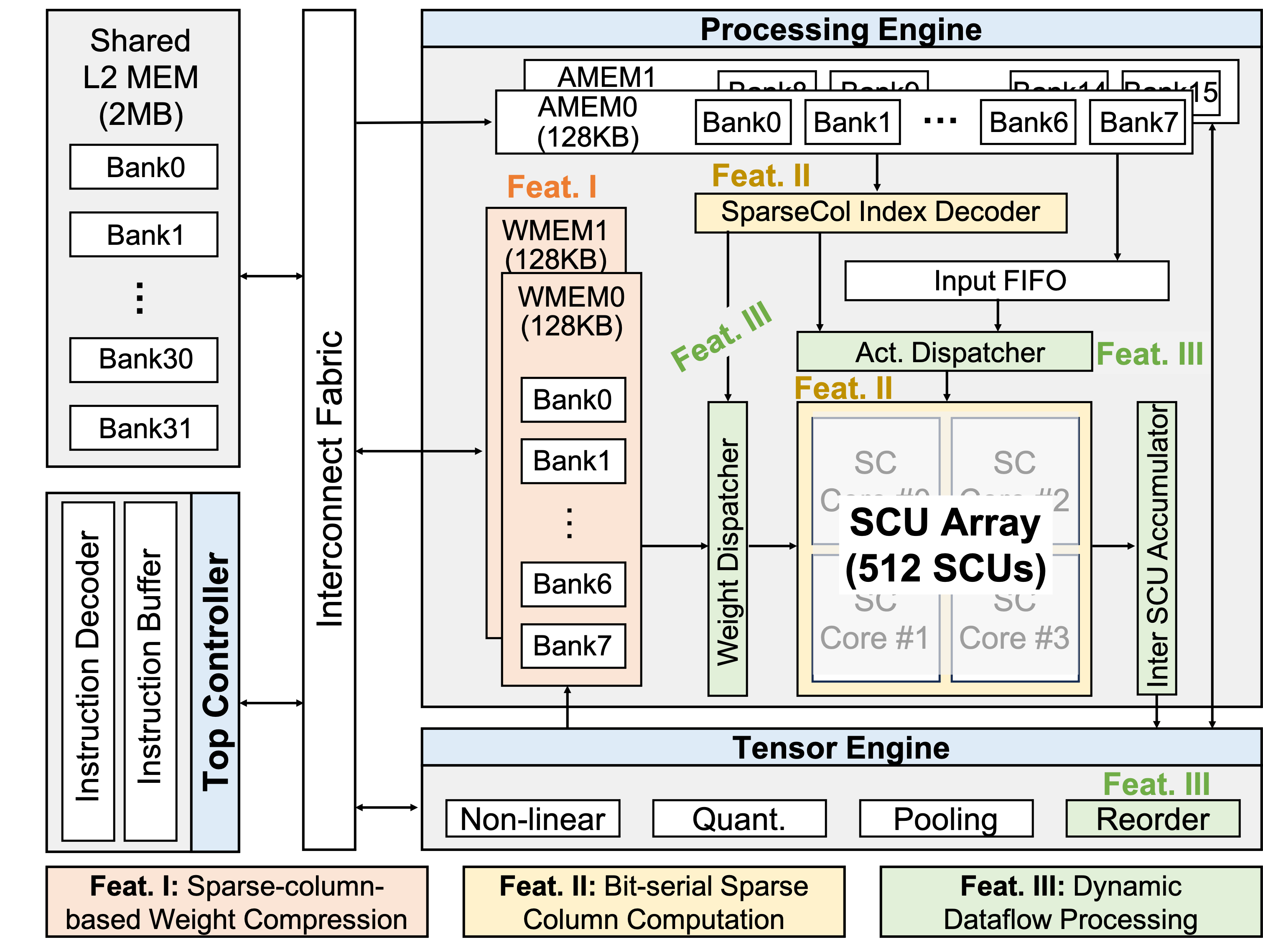}
    \caption{SparseCol system architecture overview.}
    \label{fig:fig4}
    \vspace{-0.2cm}
\end{figure}

\section{\mv{SparseCol} System Architecture}
\label{sec:system_architecture}

This section provides a high-level architectural overview of the \textit{SparseCol} system. Fig.~\ref{fig:fig4} depicts the full \textit{SparseCol} architecture: a top controller, two computing engines, a hierarchical distributed memory subsystem for operand storage, and interconnection infrastructure that completes the overall system design. The subsequent discussion will address the control system design \mv{and} memory subsystem organization and optimization. 
Detailed architectures of the two engines are presented in Sec~\ref{sec:core1} and Sec~\ref{sec:core2}.

\subsection{Top controller architecture}
\begin{figure}[t]
    \centering
\includegraphics[width=1\columnwidth]{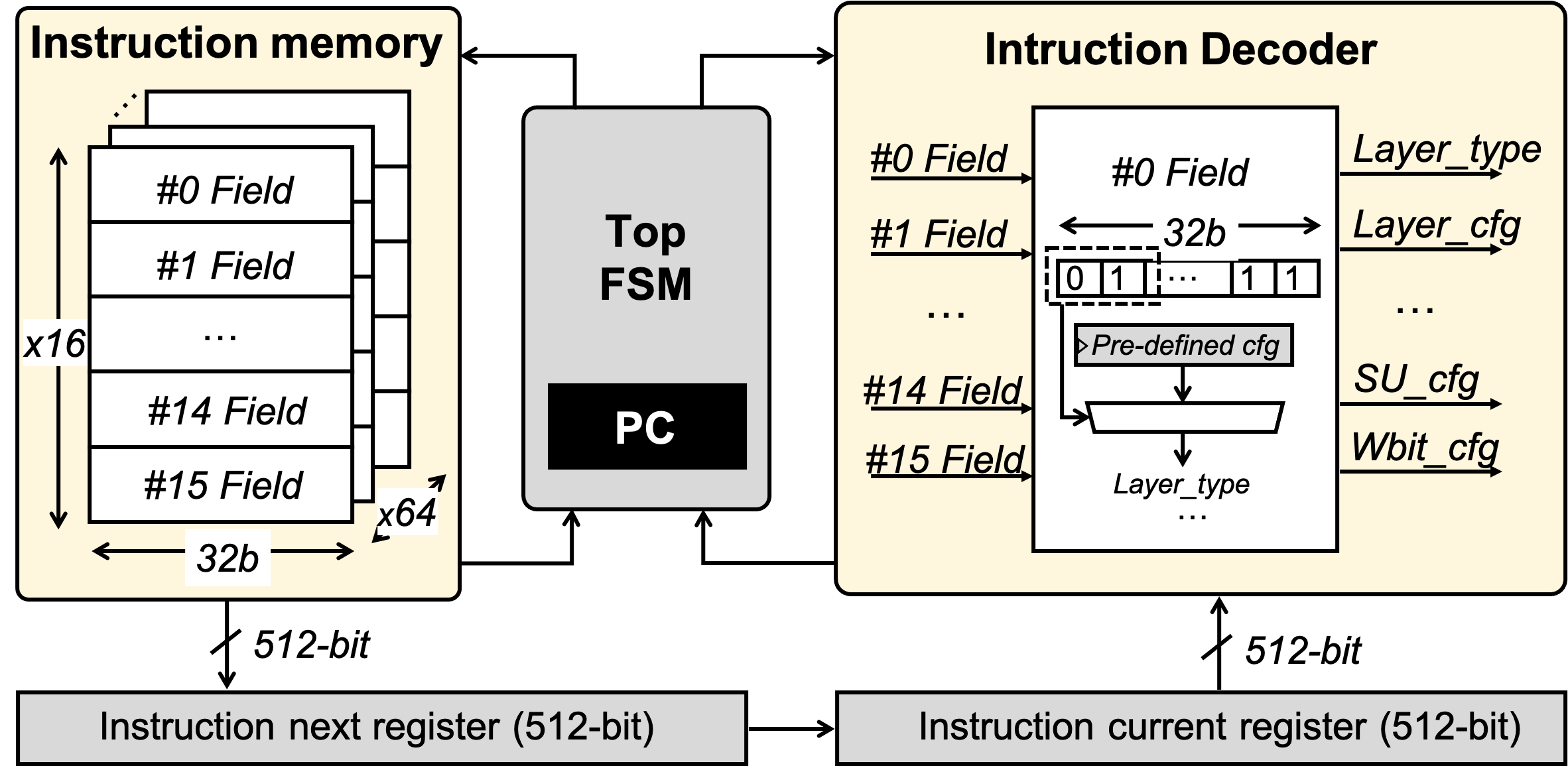}
    \caption{Top controller architecture with instruction buffer and instruction decoder.}
    \label{fig:fig5}
    \vspace{-0.2cm}
\end{figure}

\mv{The \textit{SparseCol} processor is instruction set programmable.} Fig.~\ref{fig:fig5} illustrates the top controller micro-architecture consisting of an instruction memory subsystem and an instruction decoder unit. The instruction memory contains \mv{for each instruction 16 fields of 32-bit each and has space for 64 instructions,} 
providing a total instruction space of $512 \times 64$ bits. A Top FSM (finite state machine) with an integrated PC (program counter) controls instruction fetching and sequencing.
The instruction decoder receives the 512-bit instruction stream and parses individual fields to extract configuration parameters, such as layer type identification (can also function as independent on-chip memory space access workloads), layer configuration settings, and so on. 
The system employs a two-stage instruction pipeline with dedicated 512-bit registers: an instruction-next register for prefetching upcoming instructions and an instruction-current register for holding the currently executing instruction. This pipelined approach enables a continuous instruction flow while the decoder processes complex multi-field instructions. 
The entire architectural framework maintains configurability through this instruction-set programmability under the centralized Top Controller, providing system-wide coordination and operational management.



\subsection{Memory system and optimizations}
\label{sec:mem}
\begin{figure}[t]
    \centering
\includegraphics[width=1\columnwidth]{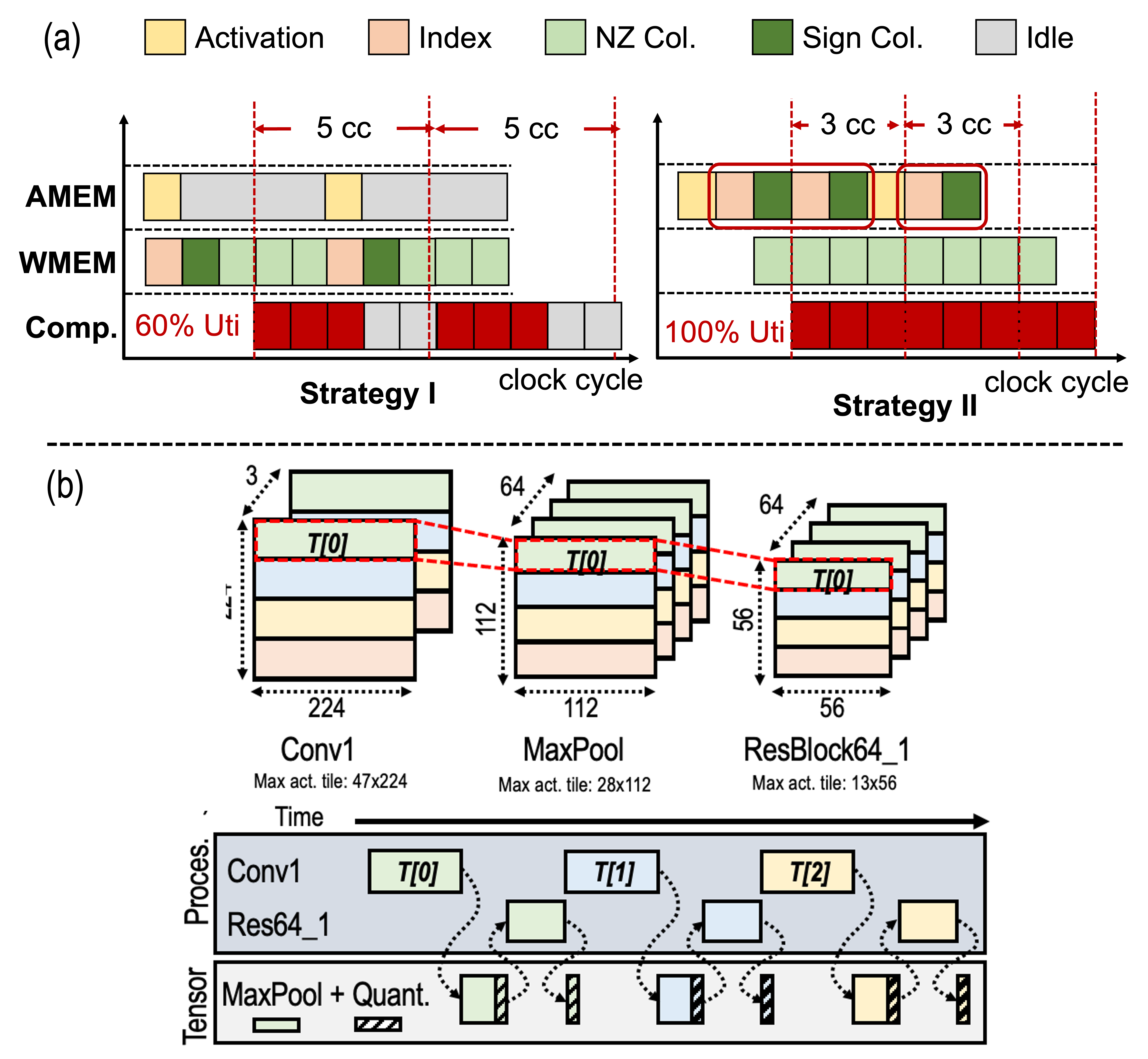}
    \caption{Memory subsystem two optimizations: (a) The L1 on-chip memory (AMEM and WMEM) organization and access strategies comparisons. (b) The L1 AMEM storage requirement optimization with layer fusion technique.}
    \label{fig:fig13-1}
    \vspace{-0.2cm}
\end{figure}

\textit{SparseCol} implements a three-tier hierarchical memory architecture comprising a global L2 SRAM, dedicated L1 SRAM scratchpads, and distributed register-based local storage (L0).
The L2 SRAM serves as the primary scratchpad with 2MB capacity for storing inputs, outputs, weights, and indices. This memory is strategically partitioned across 32 independent banks, each providing 4-byte read/write bandwidth to maximize parallel access throughput. Two specialized 256KB L1 memory modules: WMEM for compressed weights, and AMEM for activations, compressed weight sign, and weight index information; 
each feature 16 SRAM banks supporting a  bandwidth of up to 1024-bit per clock cycle.
Both L2 and L1 SRAM modules can be flexibly accessed for read and write operations from external interfaces, enabling comprehensive system debugging and verification capabilities. Local data reuse is enhanced through distributed L0 registers integrated within the processing engines. The \mv{specific} register-level organization of these subsystems is \mv{detailed} in the dedicated \ms{processing} engine in Sec~\ref{sec:core1}. In addition, the memory efficiency can be further enhanced through two optimizations, as shown in Fig.~\ref{fig:fig13-1}.

\subsubsection{Memory reorganization and prefetch}
In Fig.~\ref{fig:fig13-1}(a), 
the strategy I shows \mv{the consequence of storing the weight} index
and sign non-zero bits 
in the weight memory (WMEM), \mv{together with the NZ columns}. Yet, this creates an inefficiency because activations need to be reused across multiple cycles when computing NZCs that belong to the same weight group.
This reuse pattern means the AMEM bandwidth is underutilized but the WMEM bandwidth is insufficient, resulting in significant idle computing time across the system.
The strategy II demonstrates the optimized approach \mv{which prefetches both the weight index and sign information}, completing the same operations in just 3 clock cycles. The key point involves relocating the index and nonzero sign bits from WMEM to AMEM. This reorganization allows the system to preload the index and sign bits in advance by exploiting the non-used AMEM bandwidth, effectively avoiding memory access stalls. As a result, AMEM memory space undergoes adaptive partitioning according to data (indices, sign, activation) size requirements as shown in Fig.~\ref{fig:fig6}. This configurability is supported by the instruction set that enables arbitrary data placement and type assignment throughout the AMEM address space.
We examine the latency reduction benefits \mv{of the memory layout optimization} for BERT-Base. 
While applying bit-column sparsity with Bit-Flip to BERT-Base increases the compression ratio, the latency reduction remains limited due to massive memory stalls ($1.3\times$ reduction compared to dense computation baseline). Only when combined with the proposed prefetch optimization latency decreases substantially from  $1.3\times$ to $1.8\times$. 

\subsubsection{Layer Fusion}
To reduce L1 AMEM requirements, we implement layer fusion across processing and tensor engines. The core principle of layer fusion~\cite{alwani2016fused} involves initiating execution of subsequent layers immediately upon partial completion of preceding layers, rather than waiting for complete layer processing. For ResNet18, we fuse three consecutive layers: the initial convolutional layer, the max pooling layer, and the first layer of ResBlock64 (Fig.~\ref{fig:fig13-1}(b)).
The output feature map from Conv1 has dimensions 112×112×64, requiring 784 KB of L1 memory space, significantly exceeding our 256 KB L1 AMEM capacity. Without layer fusion, this would trigger expensive L2 memory transfers to hold intermediate results. Instead, we partition the activation tensor into smaller tiles and process them sequentially. However, smaller tiles reduce memory requirements at the cost of computational overhead due to redundant processing of overlapping regions between adjacent tiles~\cite{9570718}.
Through experimental analysis, processing 56×13-pixel activation tiles in ResBlock64\_1 achieves optimal performance, balancing memory reduction with computational overhead. This tiling strategy enables immediate reuse of intermediate data within L1 memory, achieving a $3.1\times$ reduction in L1 memory space requirements. 

\begin{figure}[tb]
\centering
\includegraphics[width=1\linewidth]{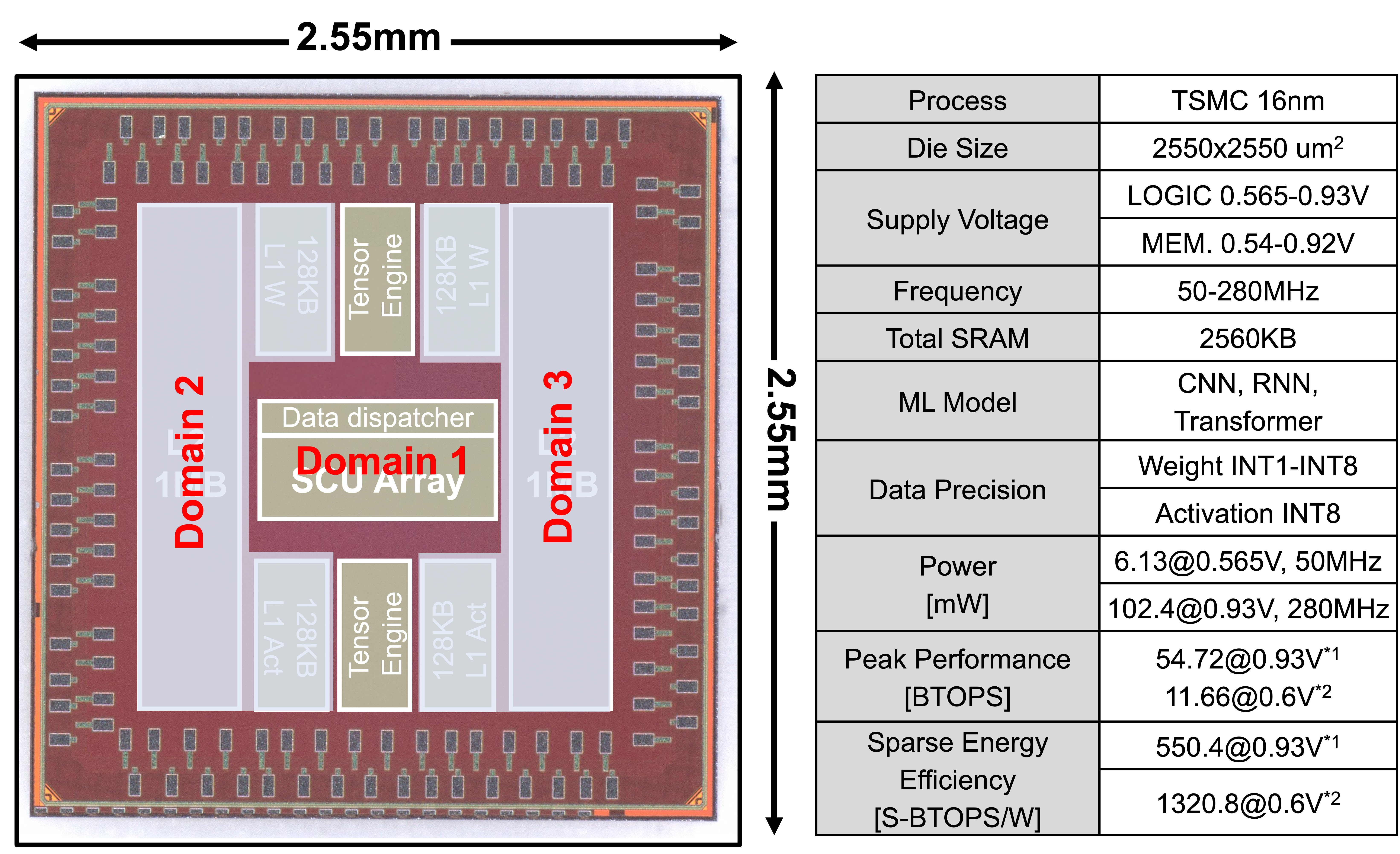}
\caption{Left: Chip power domain design, including three power domains: one logic domain and two memory domains. \ms{Right: Chip specification summary.}} 
\label{fig:domain}   
\vspace{-0.3cm}
\end{figure}
\subsection{Physical Design}
The custom \textit{SparseCol} with processing and tensor engines was manufactured using TSMC's 16nm CMOS fabrication process with an area of 6.5 $mm^2$. Fig.~\ref{fig:domain} shows an annotated diagram of the floorplan \ms{and the chip spec summary}. 


\ms{The chip timing signoff analysis was conducted at 0.72V under the slow-slow (SS) process corner to verify setup time constraints and at 0.88V under the fast-fast (FF) process corner to validate hold time requirements. \textit{SparseCol} achieves a maximum operating frequency of 280MHz at supply voltages of 0.93V for logic and 0.92V for memory. The clock distribution network is driven directly by an external clock generator without intermediate PLL or FLL circuitry. Currently, the critical timing path traverses from the activation memory through the data dispatcher to the index parser, which constrains the maximum achievable frequency. Further performance enhancement is feasible through the insertion of additional pipeline stages; since all data is prefetched, the introduction of these pipeline registers would not degrade computational throughput. \textit{SparseCol} employs a three-domain power architecture, strategically partitioned to optimize energy efficiency across diverse operational scenarios. The primary domain encompasses the core logic components, while two independent power domains are allocated for the on-chip memory subsystem, with each domain housing half of the memory banks at each level of the memory hierarchy. This partitioned architecture enables selective power gating of inactive memory domains during specialized operational modes, yielding significant energy savings for workloads with reduced memory footprint requirements.}
\section{MEASUREMENT RESULTS}
\label{sec:measurements}

In this section, the measurement results regarding the \textit{SparseCol}'s power consumption, energy efficiency, and overall performance characteristics are presented.

\subsection{Chip characteristics v.s. voltage, frequency, and power} 
\begin{figure}[tb]
\centering
\includegraphics[width=1\linewidth]{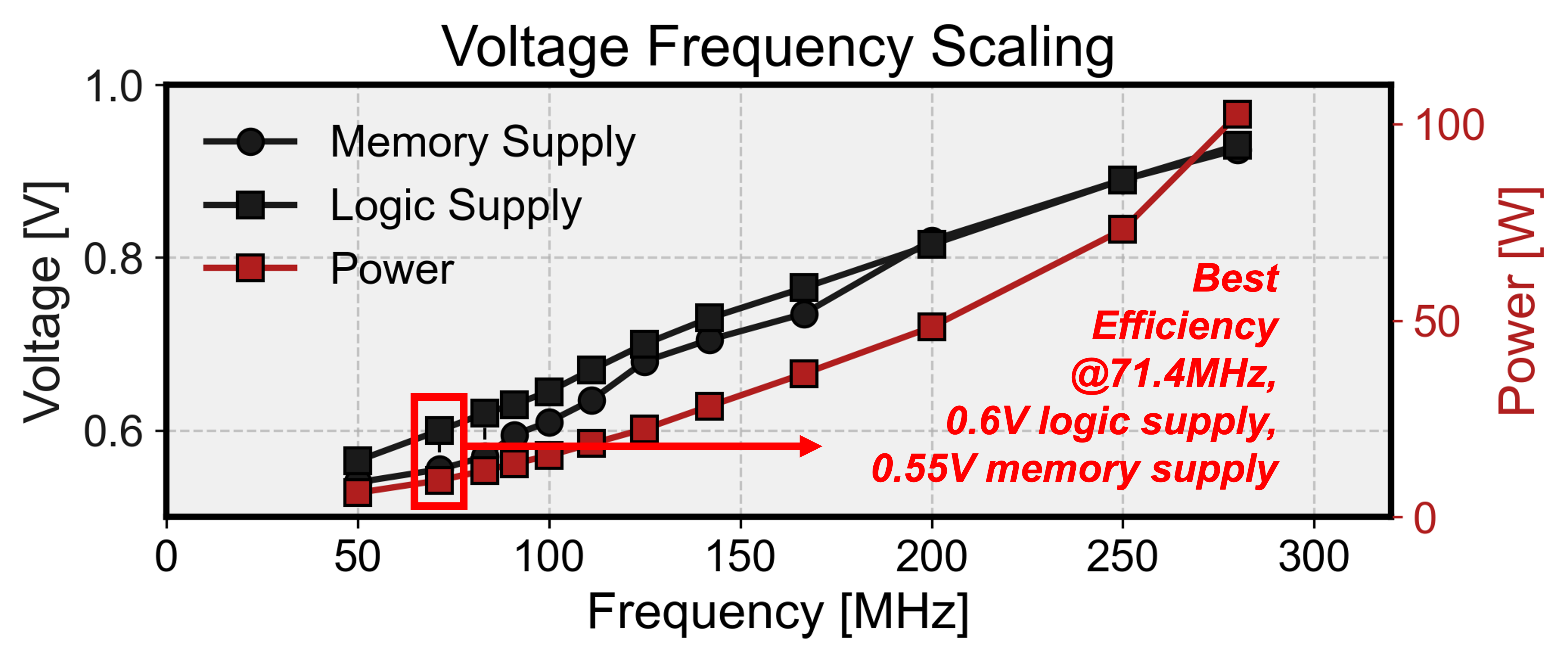}
\caption{\ms{SparseCol Chip frequency, voltage, and power scaling characteristic}} 
\label{fig:chip} 
\vspace{-0.2cm}
\end{figure}

An initial power assessment was conducted employing a single CNN layer configuration comprising 128 input channels, 128 output channels, and a 3×3 filter kernel with a random Gaussian distribution. This particular layer architecture was selected for peak performance evaluation due to: (1) the prevalence of 3×3 filter kernels in contemporary DNN frameworks. 
(2) maximizing utilization of the PE array. (3) accommodating the capacity limitations of the accelerator's L1 memory structures. 
The relationship between the clock frequency and power \mv{consumption} was systematically examined through \ms{statically adjusting the both logic components, memory subsystems (while only turning on one memory domain) and clock frequency across different operating points while maintaining correct workload execution}, as illustrated in Fig.~\ref{fig:chip}. The optimal efficiency point occurs at 71.4 MHz with 0.6V logic supply and 0.55V memory supply, while the best performance point is at 280 MHz with 0.93V logic supply and 0.9V memory supply.

\subsection{Chip characteristics v.s. Number of \mv{zero} columns}
\begin{figure}[tb]
\centering
\includegraphics[width=1\linewidth]{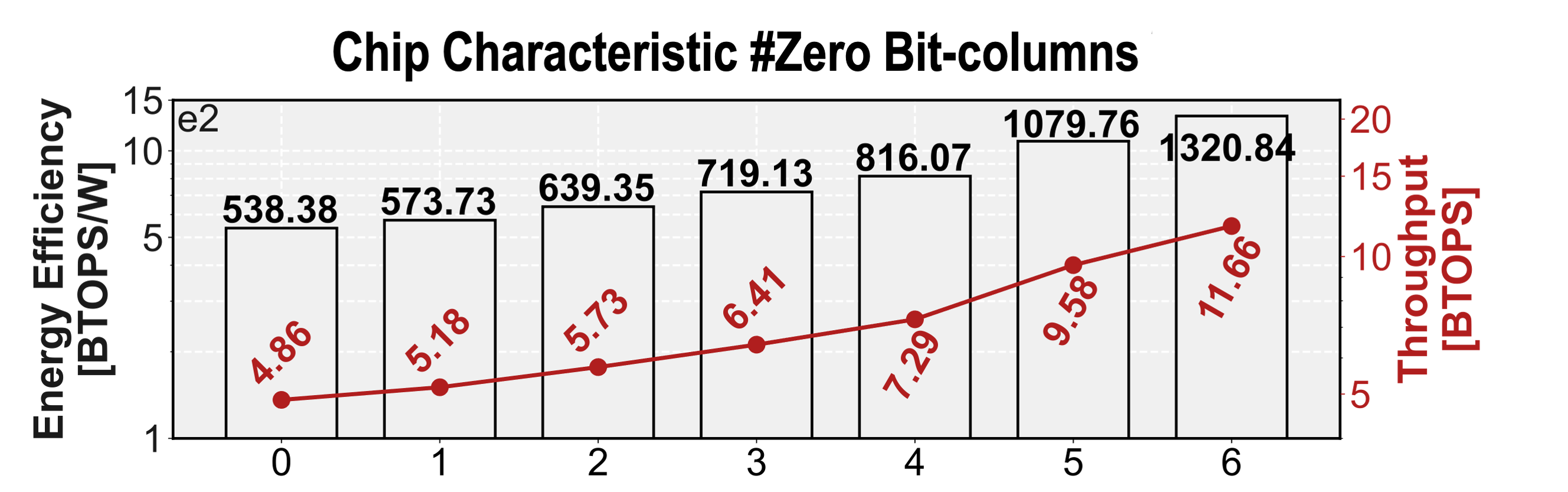}
\caption{SparseCol chip characteristics under the different number of zero columns \ms{(forced by Bit-Flip)}, Measured on 128x128x3x3 kernel with Gaussian distributed random operands @0.6V, 71.4MHz.} 
\label{fig:12} 
\end{figure}

We evaluate system performance and energy efficiency as a function of sparsity level, specifically varying the number of zero bit-columns (\#Sparse Columns). Fig.~\ref{fig:12} shows the relationship between zero bit-column count (ranging from 0 to 6; 6 zero columns is the maximum number that is reachable in some benchmarks with negligible model performance degradation. 7 zero columns violate accuracy constraints across all benchmarks) and two key metrics: energy efficiency and peak throughput.
Energy efficiency is measured in BTOPS/W, where BTOPS (Binary Tera-Operations Per Second) \cite{Abati_2023_ICCV, Baskin_2019} is calculated as $\#Wbits \times \#Abits \times TOPS$. Energy efficiency measurements are taken at the optimal operating point of 0.6V logic supply, 0.55V memory supply, and 71.4MHz frequency. Peak throughput is measured in BTOPS at the maximum performance operating point with 0.93V logic supply, 0.9V memory supply, and 280MHz frequency.

As the number of zero columns increases from 0 to 6, energy efficiency demonstrates a consistent upward trend, rising from 538.38 BTOPS/W with no zero columns to 1320.84 BTOPS/W with six zero columns, representing a $145\%$ improvement. Concurrently, computational throughput exhibits a similar positive correlation, increasing from 4.86 BTOPS to 11.66 BTOPS, a $140\%$ gain. 
Notably, the improvement in both metrics does not scale perfectly linearly with the number of sparse columns due to memory access inefficiencies that create system stalls. While computational workload decreases proportionally with each additional zero column, the memory subsystem incurs persistent overhead from imperfect control flow that is not fully optimized in current implementation.
Future work will focus on optimizing the system control logic to better approach the theoretical performance limits.


\begin{figure}[tb]
\centering
\includegraphics[width=0.9\linewidth]{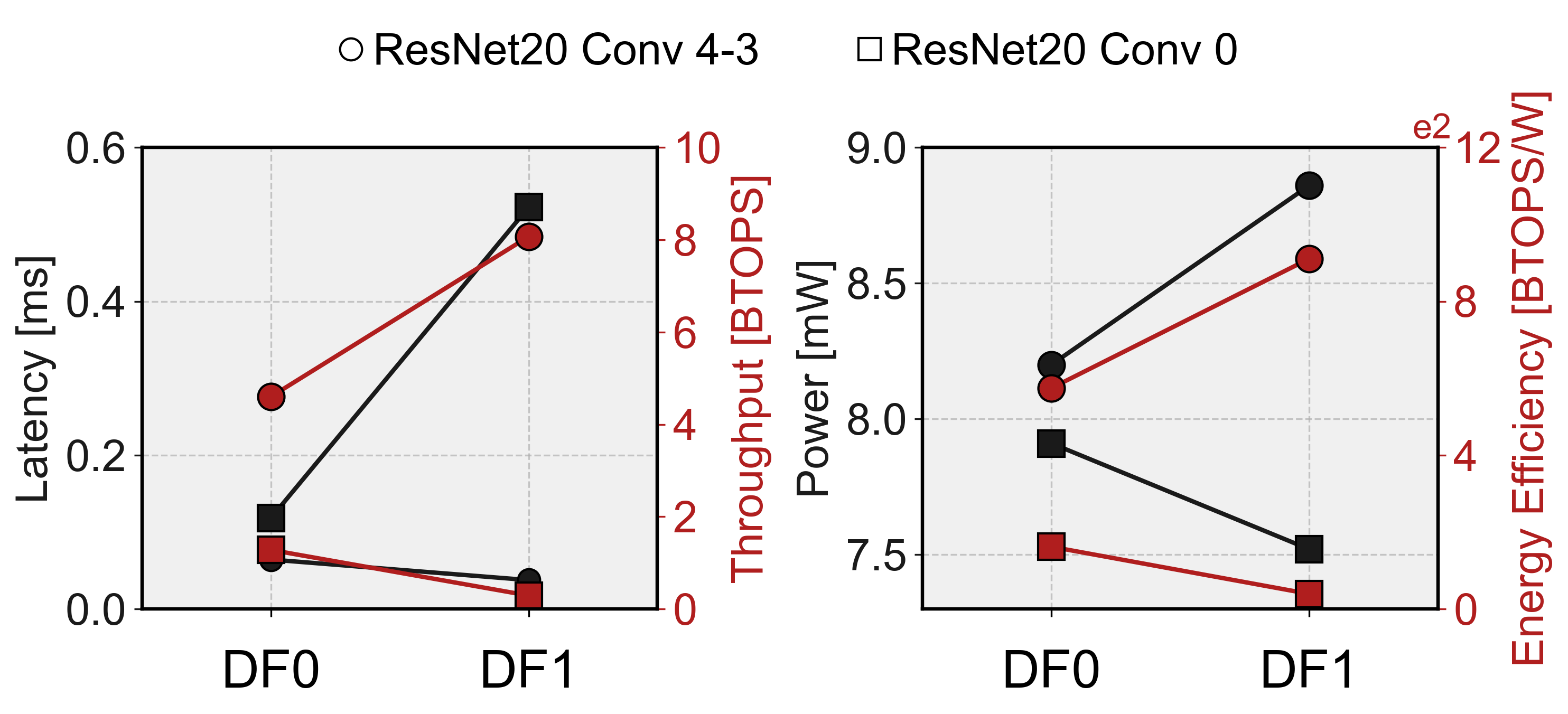}
\caption{\ms{System latency, power and energy efficiency for ResNet20 convolutional layers under DF0 and DF1.}} 
\label{fig:df1} 
\vspace{-0.4cm}
\end{figure}

\begin{figure}[tb]
\centering
\includegraphics[width=0.9\linewidth]{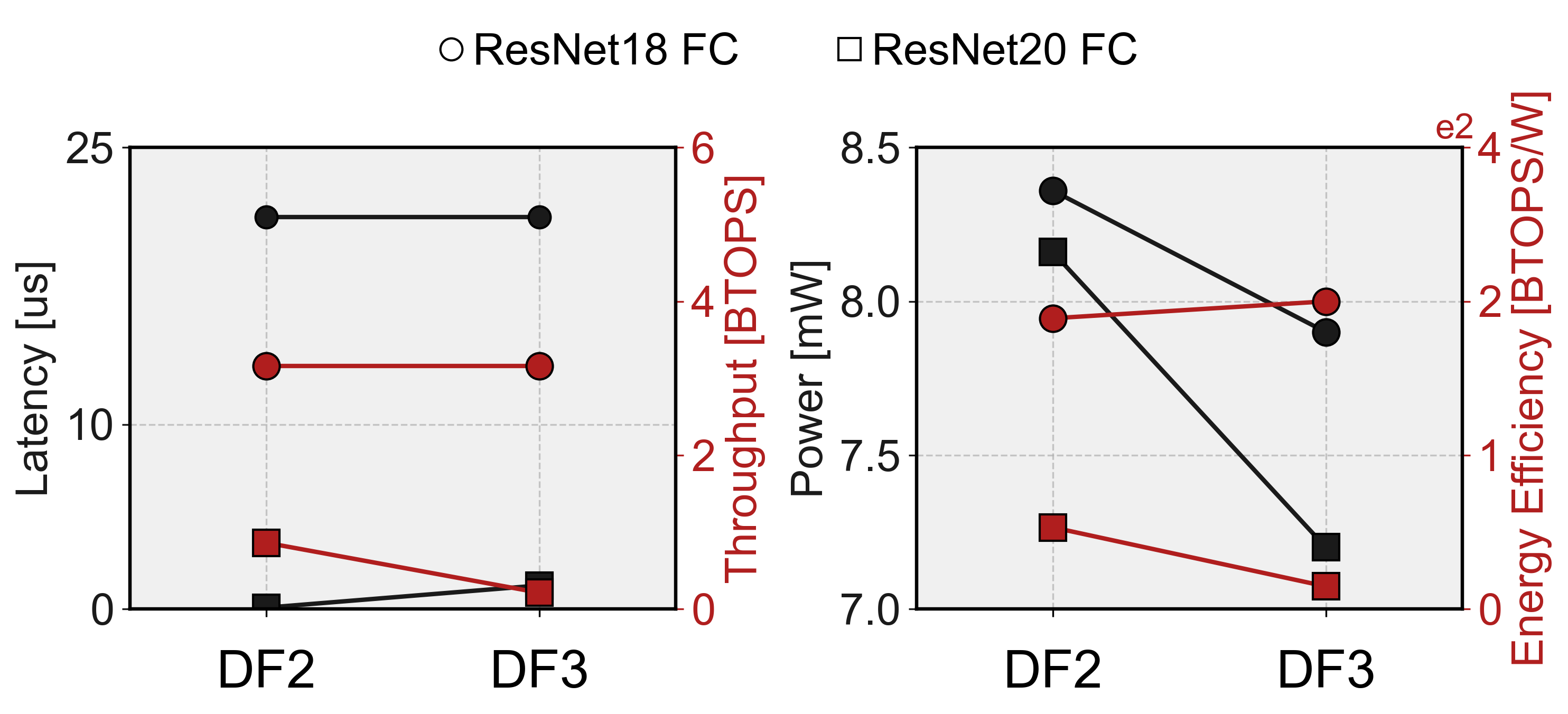}
\caption{\ms{System latency, power and energy efficiency for ResNet18 FC layer, ResNet20 FC layer under DF2 and DF3.}} 
\label{fig:df2} 
\vspace{-0.4cm}
\end{figure}
\subsection{Chip characteristics v.s. DataFlow}
\mv{Next, we} conduct the performance, power and efficiency analysis across different dataflows, demonstrating the importance of supporting dynamic dataflow. Fig.~\ref{fig:df1} and Fig.~\ref{fig:df2} show a detailed comparison across different neural network architectures (ResNet18 and ResNet20) layer types and topologies under different dataflow (DF) configurations. 

In Fig.~\ref{fig:df1}, which focuses on ResNet20 convolutional layers (Conv 4-3 and Conv 0), we observe significant variations in latency and throughput across DF0 and DF1.
Conv 0 demonstrates lower latency and higher energy efficiency when executed using DF0. This performance can be attributed to Conv 0's wide feature map, which is optimized by DF0 and enables maximal PE utilization and energy efficiency. Conversely, Conv 4-3 exhibits superior performance under DF1 due to its deep channel architecture, which is strategically optimized to enhance PE utilization.
Fig.~\ref{fig:df2} compares ResNet18's and ResNet20's fully connected (FC) layers, offering an intriguing counterpoint. In this context, the latency remains remarkably consistent across DF2 and DF3 for the ResNet18 FC layer, as both dataflows maintain equivalent PE utilization. However, DF2, characterized by larger output channel parallelism, proves less advantageous for the ResNet20 FC workload due to its narrow output channel, consequently resulting in reduced computational efficiency. 
These visualizations underscore a critical insight: the choice of dataflow is paramount in optimizing workload mapping on neural network accelerators. Different network layers and architectures respond distinctly to various dataflow strategies, highlighting the importance of tailored computational approaches. 

\begin{table*}[h!]
\centering
\renewcommand{\arraystretch}{1.2}  
\begin{threeparttable}
  \small 
\caption{Workload measurement results}
  \scriptsize
\begin{tabular}{>{\columncolor{gray!30}}r|cccc}
\hline
\rowcolor{gray!30} 
\textbf{Dataset} & \multicolumn{2}{c}{ImageNet} & MRPC & DNS Challenge ~\cite{russakovsky2015imagenet}\\ \hline
\textbf{Task} & \multicolumn{2}{c}{Image Classification} &  Sequence Classification & Speech Enhancement \\ \hline
\textbf{Network} & ResNet18~ & MobileNetV2~ & Bert-Base~\textsuperscript{*1} & CRUSE, CNN-LSTM stack \\ \hline
\textbf{Baseline FP Accuracy} & 69.75 (\%) & 71.88 (\%) & 0.889, F1 score& 2.953, PESQ score@12db SNR \\ \hline
\textbf{Baseline Int8 Accuracy~\textsuperscript{*2}} & 69.35  & 71.68  & 0.884& 2.953 \\ \hline
\textbf{Accuracy w/ Bit-Flip} & 69.12 & 71.10& 0.88 & 2.938 \\  \hline
\textbf{Average $\#$Zero Bit-column \textsuperscript{*3}} & 3.21 & 2.94 & 4.17 &5.42 \\ 
\hline
\textbf{Inference Per Second\textsuperscript{*4}} & 27.8-105.4 & 41.3-179.1  & 11.2-41.1 & 4485-16948 \\ \hline
\textbf{Performance Speedup\textsuperscript{*5}} & $\times$2.13 & $\times$1.83  & $\times$1.77 & $\times$1.98 \\ \hline
\textbf{Sparse Energy Efficiency}  & 254.4@0.93V& 231.2@0.93V & 311.1@0.93V & 42.1@0.93V \\
 \textbf{[S-BTOPS/W]\textsuperscript{*6}} & 738.6@0.6V & 649.3@0.6V & 850.1@0.6V & 127.2@0.6V  \\ \hline
\end{tabular}
\begin{tablenotes}
\item[*1]: Bert-Base model is evaluated based on a 256 sequence length.
\ms{\item[*2]: Open-source Int8 quantized networks, or - where not
available - quantized the fp32 weights into Int8 precision using
PyTorch’s common post-training quantization framework.} 
\item[*3]: \ms{The avg. \#zero bit-columns is forced by layer-wise Bit-Flip technique, reflecting the actual compressed representation stored in WMEM and considering the weight group overheads.}
\item[*4]: Excluding off-chip memory to focus on on-chip optimization. CRUSE inference focuses on LSTM layers. 
\item[*5]: Baseline: chip without bit-column sparsity related features.
\item[*6]: Including the skipped operations, BTOPS (=Binary Tera-Operations Per Second = \#Wbits*\#Abits *TOPS). 
\end{tablenotes}   
\label{table:1}
\end{threeparttable}
\end{table*}

\subsection{Chip characterization v.s. workload}

\begin{table}[tb]
\centering
\caption{Core area and Power Characteristics}
\label{tab:area_breakdown}
\resizebox{0.48\textwidth}{!}{%
\begin{tabular}{@{}llcc@{}}
\toprule
Component               & Setup                         & Area {[}mm$^2${]}        & Power {[}mW{]}            \\ \midrule
L2 Memory & 2MB & 2.02 (43.8\%)           & 6.80 (14.0\%)           \\
L1 Memory    & 512KB                      & 0.57 (12.4\%)            & 10.1 (20.9\%)             \\
Index Decoder             & 128 8-bit Decoder                      & 0.05 (1.07\%)            & 1.80 (3.73\%)             \\
Data Dispatcher       & Act./Weight Dispatcher  & 0.22 (4.78\%)           & 3.10 (6.42\%)           \\
SCU Array            & 512 SCU                           & 1.36 (29.6\%)           & 22.6 (46.8\%)             \\
Tensor Engine            & Pooling/Quant/Non-linear/Reorder                           & 0.37 (8.04\%)           & 1.70 (3.52\%)             \\
Instruction Memory            & 4KB RF                          & 0.01 (0.21\%)           & 0.90 (1.86\%)             \\
Controller                  & Top controller                & 0.01 (0.22\%)            & 1.30 (2.69\%)             \\ \midrule
\textbf{Total}          & \textbf{}                     & \textbf{4.61 (100\%)} & \textbf{48.3 (100\%)} \\ \bottomrule
\end{tabular}%
}
\vspace{-0.3cm}
\end{table}

\begin{table*}
    \caption{Comparison with State-of-the-Art sparse accelerators}
        \small 
    \begin{threeparttable}
    \scriptsize
\resizebox{0.98\textwidth}{!}{
    \centering
    \renewcommand{\arraystretch}{1}  
	\begin{tabular}{r|cccccc}
		\hline
         \rowcolor{gray!30}
         \cellcolor{gray!30} & \textbf{Bitlet\textsuperscript{*1}} & \textbf{DA-SignM} & \textbf{Slim-Llama} & \textbf{HUAA} & \textbf{Onyx} & \textbf{This} \\
         \rowcolor{gray!30}
         \cellcolor{gray!30}~ & HPCA'21~\cite{bitlet} & ISSCC'23~\cite{10067269}  & ISSCC'25~\cite{10904761} & ISSCC'23~\cite{du202328nm} & VLSI'24~\cite{10631383}& \textbf{work} \\
         \cline{1-7}
         \hline
         \cellcolor{gray!30} \textbf{Technology} & 28nm & 28nm & 28nm & 28nm & 12nm & \textbf{16nm} \\
         \cline{1-7}
         \hline 
         \cellcolor{gray!30} \textbf{Area [mm$^\text{2}$]}  & 1.54 & 7.75 & 20.25 & 7.81 & 23 & \textbf{6.5} \\
         \hline
         \cellcolor{gray!30} \textbf{Frequency [MHz]}  & 1000 & 55-285 & 25-200& 100-500 & 500-980 & \textbf{50-280} \\
         \cline{1-7}
         \hline     
         \cellcolor{gray!30} \textbf{Voltage [V]}  & - & 0.65-0.9 & 0.58-1.0 & 0.66-1.3 & 0.6-1.0 & \textbf{0.565-0.93} \\
         \cline{1-7}
         \hline
         \cellcolor{gray!30} \textbf{Number of MACs \textsuperscript{*2}}  & 96 & 1024 & 512& 1024 & 1536 & \textbf{512} \\
         \hline
         \cellcolor{gray!30} \textbf{On-Chip Mem. [KB]} & - & 1074 & 500  & 1120 & 4500 & \textbf{2560} \\
         \hline
         \cellcolor{gray!30} \textbf{Sparsity Support} &  W bit & W bit & W bit & None & W/A value & \textbf{W bit} \\
         \cline{1-7}
         \hline
         \cellcolor{gray!30} \textbf{Flexible Dataflow} & None & None & None & Yes & None &  \textbf{Yes} \\
         \cline{1-7}
         \hline
         \cellcolor{gray!30} \textbf{Power [mW]} & - & 6.6-179.4 & 4.69-82.07 & 17-174 & - & \textbf{6.13-102.4}  \\
         \cline{1-7}
         \hline
         \cellcolor{gray!30} \textbf{Peak Performance} & \multirow{2}{*}{47.66} & 27.97@0.9V &  \multirow{2}{*}{165.58@1.0V}  & \multirow{2}{*}{-} &  \multirow{2}{*}{73.09}  & \textbf{{54.72}@0.93V}\textsuperscript{*4} \\
         \cellcolor{gray!30} \textbf{[BTOPS]\textsuperscript{*3}}  & ~& 5.38@0.65V & ~ & ~ & ~ & \textbf{{11.66}@0.6V}\textsuperscript{*5} \\
         \hline
         \cellcolor{gray!30} \textbf{ Effective Peak Energy } & \multirow{2}{*}{85.12} & \multirow{2}{*}{-} &  \multirow{2}{*}{980@0.58V}  & \multirow{2}{*}{480-716.8} & \multirow{2}{*}{-} &  \textbf{190.71@0.93V}\textsuperscript{*6}  \\
         \cellcolor{gray!30} \textbf{Effi. [E-BTOPS/W]\textsuperscript{*3}}  & ~& ~ & ~ & ~ & ~ &\textbf{645.12@0.6V}\textsuperscript{*7}  \\
         \hline
         \cellcolor{gray!30} \textbf{Sparse Peak Energy}& \multirow{2}{*}{-}  & 268.8@0.9V & \multirow{2}{*}{2388.96@0.58V} & \multirow{2}{*}{-} &193.53\textsuperscript{*8} &\textbf{550.4@0.93V}\textsuperscript{*4}   \\
         \cellcolor{gray!30} \textbf{Effi. [S-BTOPS/W]\textsuperscript{*3}} & ~& 517.7@0.65V & ~  & ~ & @0.66V, 500MHz & \textbf{1320.8@0.6V}\textsuperscript{*5}  \\
         \hline
         \cellcolor{gray!30} \textbf{Area Efficiency} & \multirow{2}{*}{30.95}& 3.61@0.9V & \multirow{2}{*}{8.27@1.0V}  & \multirow{2}{*}{-} & \multirow{2}{*}{1.59} & \textbf{8.41@0.93V}\textsuperscript{*4}   \\
         \cellcolor{gray!30} \textbf{[BTOPS/mm$^\text{2}$]\textsuperscript{*3}} & ~& 0.69@0.65V & ~  & ~ & ~ & \textbf{1.79@0.6V}\textsuperscript{*5} \\
         \hline
	\end{tabular}
    }
    \begin{tablenotes}
    \item *1: Simulated architecture. *2: The number of MACs is normalized the 8x8 MACs. \\
*3: One operation (OP) is one mult. or one add. \ms{For fair comparison, all SOTA results are converted using BTOPS = BitsW × BitsA × TOPS, where bit-widths represent the maximum precision supported by each accelerator. All conversions preserve the original measurement conditions (voltage, frequency, workload). Data not available in the original publication are indicated by "-" in the table.} \\
*4: At the highest performance point. 0.93V for logic, 0.92V for mem., 280MHz, \ms{50\% activation sparsity and weight dense bit-column sparsity (15\% irregular bit level sparsity)}. *5: At the highest efficiency point 0.6V for logic, 0.55V for mem., 71.4MHz, \ms{50\% activation sparsity and weight 75\% bit-column sparsity (6 zero columns, 93\% irregular bit level sparsity)}. \\
*6 *7: Peak EE excluding skipped sparse computations w/o Bit-Flip. \\
*8: Reported BTOPS/W was not clearly declared to E-BTOPS/W or S-BTOPS/W (including skipped sparse computations), used as S-BTOPS/W as sparsity is exploited.
\end{tablenotes}   
    \end{threeparttable}
       \label{table:sota}
\end{table*}

\textit{SparseCol} is evaluated across diverse application domains to demonstrate its versatility and effectiveness. Table~\ref{table:1} summarizes measurement results across four representative benchmarks: ImageNet image classification using ResNet18 and MobileNetV2, MRPC sequence classification with BERT-Base, and DNS Challenge speech enhancement using CRUSE CNN-LSTM stack.
Our Bit-Flip post-processing technique achieves minimal accuracy degradation across all benchmarks. Compared to INT8 baselines, ImageNet classification experiences accuracy drops of only 0.33\% (ResNet18) and 0.81\% (MobileNetV2). Similarly, BERT-Base on MRPC shows a negligible 0.5\% F1 score reduction, while the speech enhancement model exhibits only a 0.5\% decrease in PESQ score at 12dB SNR.
Bit-column sparsity varies significantly across workloads, with average zero bit-columns ranging from 2.94 (MobileNetV2) to 5.42 (CRUSE). However, performance improvements do not scale linearly with bit-column sparsity levels due to workload architectural characteristics. ResNet18 achieves the highest speedup ($2.13\times$) by leveraging both bit-column sparsity and dynamic dataflow optimizations. In contrast, BERT-Base realizes more modest gains due to limited layer diversity that restricts dynamic dataflow benefits. Although the CRUSE model exhibits the highest compression ratio from abundant zero columns, dense layer computations with suboptimal PE utilization limit the overall speedup to $1.98\times$.
Energy efficiency improvements are substantial across all benchmarks, with the sparse accelerator achieving 231.2-850.1 BTOPS/W at optimal energy operating points (0.6V logic, 0.55V memory). These measurements include all computational overhead, providing realistic efficiency.

\ms{\subsection{Area and Power Breakdown}
\label{sec:e6}
The \emph{SparseCol} architecture implemented in 16nm FinFET technology 
occupies 6.5 mm² total area, \ms{with 4.6 mm² dedicated to the core logic, 1.9 mm² specialized to the IO peripherals.} The chip consumes 48.3 mW when running a $128\times128\times3\times3$ kernel with Gaussian distributed random operands at 238MHz, 0.8V.
Table~\ref{tab:area_breakdown} lists the breakdown of core area and on-chip power consumption of \emph{SparseCol}. The total 2560KB on-chip SRAM accounts for the most significant portion of the area, occupying $56.2\%$ of the core area. On the other hand, the PE array consumes the largest portion of power, constituting $46.8\%$ of the total power, while imposing a $29.6\%$ core area cost.
The Data Dispatcher, responsible for supporting the dynamic dataflow, requires high flexibility to serve all inputs of the PE array. This flexibility comes at the cost of $4.8\%$ and $6.4\%$ of the total area and power, respectively. In addition, the structured bit-column sparsity approach significantly simplifies hardware complexity by exploiting regular sparsity patterns, thereby requiring minimal overhead for index decoding. Specifically, in terms of area overhead, the index decoder with 128 parallel 8-bit decoder instances amounts to approximately 3.6\% of the total SCU array area. Regarding power consumption, the decoder array consumes approximately 1.8 mW, which accounts for 3.7\% of the total power.
}

\subsection{Sota comparison}
\label{sec:sota}
Table~\ref{table:sota} summarizes the state-of-the-art NN processor specifications. 
\emph{SparseCol} is  among the first chips combining optimized dataflow flexibility with advanced sparsity handling. Hence, it is capable of reducing memory accesses and skipping redundant computations simultaneously, while maintaining high PE utilization throughout the network.
This results in \emph{SparseCol} achieving the best energy efficiency across all benchmarks, even compared to a ideal simulated architecture. \emph{SparseCol} also shows its superiority in terms of area efficiency, which outperforms all taped out chips.   
In contrast to prior methodologies that leveraged bit-level sparsity  \cite{10904761, delmas2019bit, bitlet, sharify2019laconic, albericio2017bit, yang2021fusekna} or value-level sparsity \cite{li2022ristretto, 10071080, gondimalla2019sparten, parashar2017scnn} for eliminating redundant operations, or employed computationally expensive techniques to compress irregular zero-patterns in weights and activations, \emph{SparseCol} introduces a novel approach by exploiting structured bit-column sparsity patterns through hardware-optimized sign-magnitude weight representation. Furthermore, while existing approaches~\cite{zhao2020bitpruner, li2022bitcluster} require retraining procedures for workload balancing, and others~\cite{keller202395, xiao2023smoothquant} achieve only marginal performance improvements due to inherent compression-accuracy trade-offs, \emph{SparseCol} employs a training-free "one-shot" Bit-Flip technique that delivers substantial performance gains without sacrificing model accuracy. Therefore, at the optimal efficiency (0.6V, 71.4MHz), \emph{SparseCol} achieves 645 E-BTOPS/W (effective energy efficiency, excluding skipped sparse computations w/o Bit-Flip.), a $2.43\times$ improvement over current solutions, and $6.82\times$ higher sparse efficiency (S-BTOPS/W) versus a more advanced 12nm digital sparse processor.

\section{Conclusion}



We implemented, \emph{SparseCol}, a processor introducing an innovative approach called bit-column-serial computation. \emph{SparseCol} efficiently reduces redundant computation and memory access by leveraging structured bit-level sparsity in sign-magnitude represented weights, 
Combining dynamic dataflow and the Bit-Flip method, \emph{SparseCol} achieved an impressive performance increase with minimal accuracy drop (less than $1\%$ across diverse workloads and tasks). Specifically, it brings $645.12$ S-BTOPS peak performance and $1320.8$ S-BTOPS/W peak energy efficiency, outperforming a 12 nm  advanced sparse accelerator by $2.8\times$ \mv{in terms of} performance and $6.8\times$ in terms of efficiency.
\section{Acknowledgment}
This project has been partly funded by the European Research Council (ERC) under grant agreement No. 101088865, Vlaio, NXP and the Flanders AI Research Program.
\vspace{-0.5cm}
\begin{IEEEbiography}
[{\includegraphics[width=1in,height=1.25in,clip,keepaspectratio]{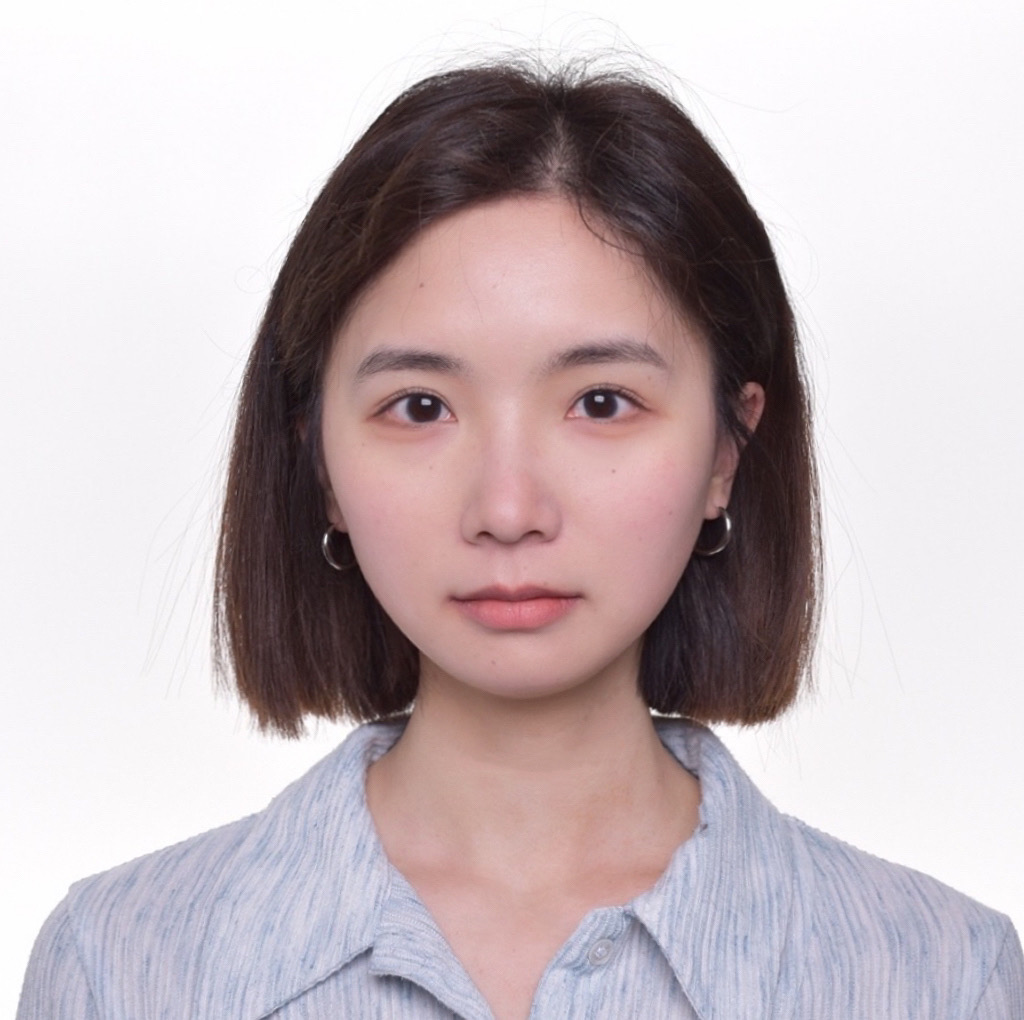}}]%
{Man Shi} is a postdoc researcher in the BWRC/SLICE lab of University of California, Berkeley. She received a Ph.D. from MICAS Laboratories at KU Leuven in 2025, an M.Sc. in Electronic Science and Technology from Tsinghua University, China (2020), and a B.Eng in Electronic Science and Technology from ShanDong University, China (2017).  Her Research Interests include AI/ML accelerators, hardware-software co-design, and Low-power chip design.
\end{IEEEbiography}

\vspace{-0.5cm}
\begin{IEEEbiography}
[{\includegraphics[width=1in,height=1.25in,clip,keepaspectratio]{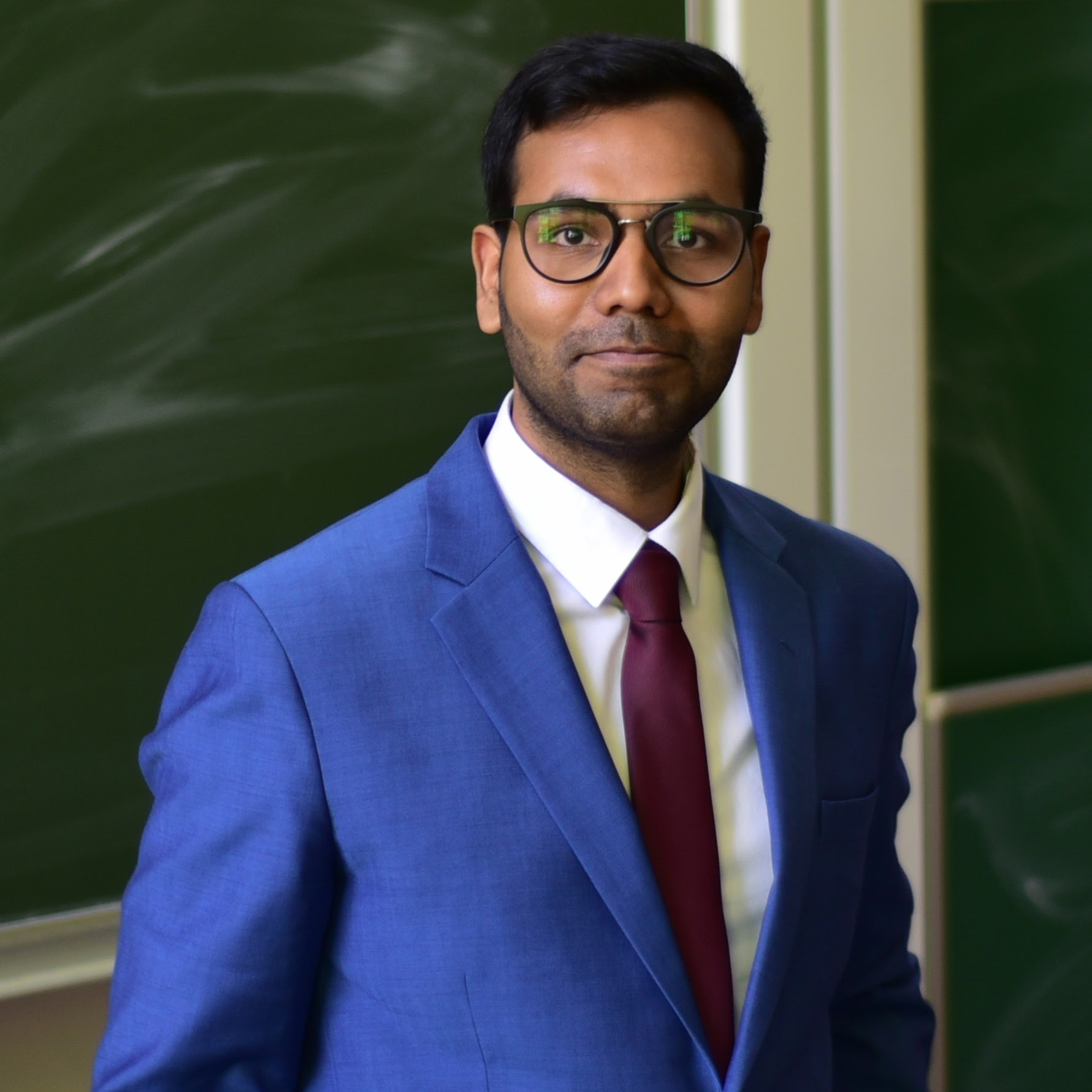}}]%
{Vikram Jain} is an incoming Assistant Professor in the Department of Electrical and Computer Engineering at Purdue University. He is currently a Principal Engineer at TSMC, San Jose. He was a Postdoctoral researcher and Lecturer at the SLICE and BWRC labs at the University of California, Berkeley. His research focuses on heterogeneous integration and chiplet architectures (2.5D and 3D) for emerging high-performance computing and AI applications. Vikram earned his Ph.D. from the MICAS laboratories at KU Leuven, Belgium. He has published numerous papers, workshops, and posters in leading conferences and journals, including ISSCC, JSSC, the Symposium on VLSI Technology and Circuits (VLSI), MICRO, HPCA, ISLPED, DAC, ISCAS, DATE, TCAS-I, TVLSI, and TC. Vikram received the Solid-State Circuits Society (SSCS) Predoctoral Achievement Award for his contributions to embedded machine-learning hardware design during 2022-2023.
\end{IEEEbiography}

\vspace{-0.5cm}
\begin{IEEEbiography}
[{\includegraphics[width=1in,height=1.25in,clip,keepaspectratio]{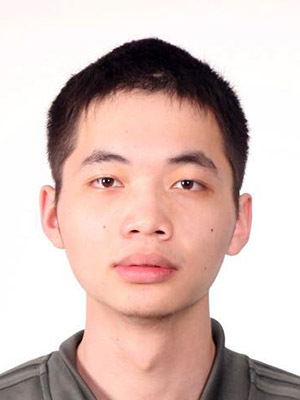}}]%
{Weijie Jiang} was born in Hubei, China, in 1996. He pursued his undergraduate studies in microelectronics at Shanghai Jiao Tong University from 2014 to 2018, where he obtained his bachelor’s degree. In 2018, he moved to Belgium to continue his education in electronic engineering at KU Leuven (Katholieke Universiteit Leuven), earning his master’s degree in 2020. He subsequently began his PhD at KU Leuven under the supervision of Prof. Wim Dehaene, focusing on SRAM and in-memory compute circuit design. 
\end{IEEEbiography}
\vspace{-0.5cm}
\begin{IEEEbiography}
[{\includegraphics[width=1in,height=1.25in,clip,keepaspectratio]{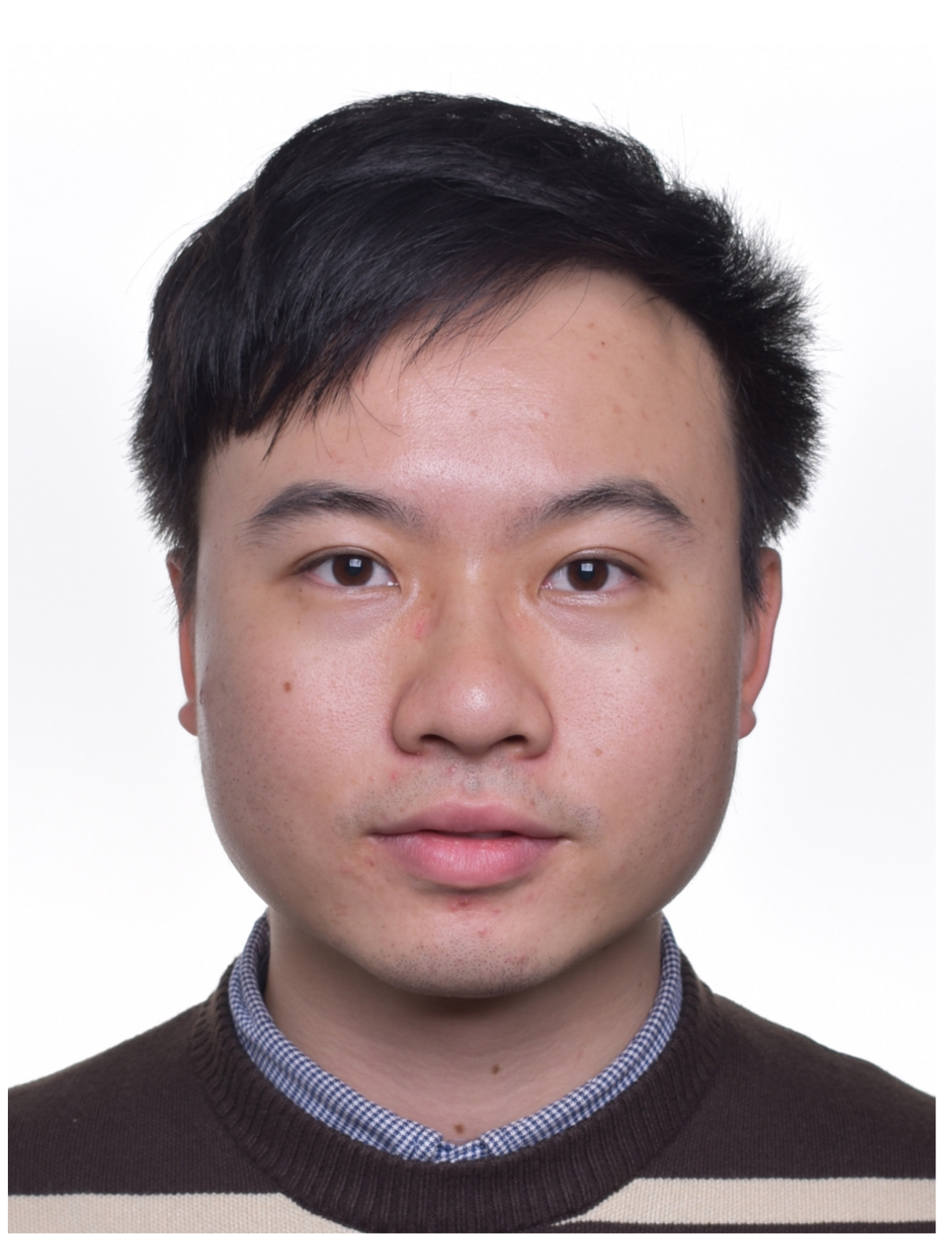}}]%
{Chao Fang} received the B.E. degree at Tianjin University, Tianjin, China, in 2019, and the Ph.D. degree at Nanjing University, Nanjing, China, in 2025. He is currently a postdoctoral researcher at KU Leuven, Belgium. He was an assistant researcher with Shanghai Qi Zhi Institute, Shanghai, China, in 2025. He served as a visiting Ph.D. scholar in the MICAS research group at KU Leuven, Belgium, in 2024. His current research interests focus on algorithm-hardware co-optimization for deep neural networks (DNNs), with particular emphasis on efficient hardware architectures for large language models (LLMs), sparse DNN computing optimization, and RISC-V processor integration with DNN accelerators. He was a recipient of the 2025 IEEE Trans. on VLSI Systems Best Paper Award.
\end{IEEEbiography}

\vspace{-0.5cm}
\begin{IEEEbiography}
[{\includegraphics[width=1in,height=1.25in,clip,keepaspectratio]{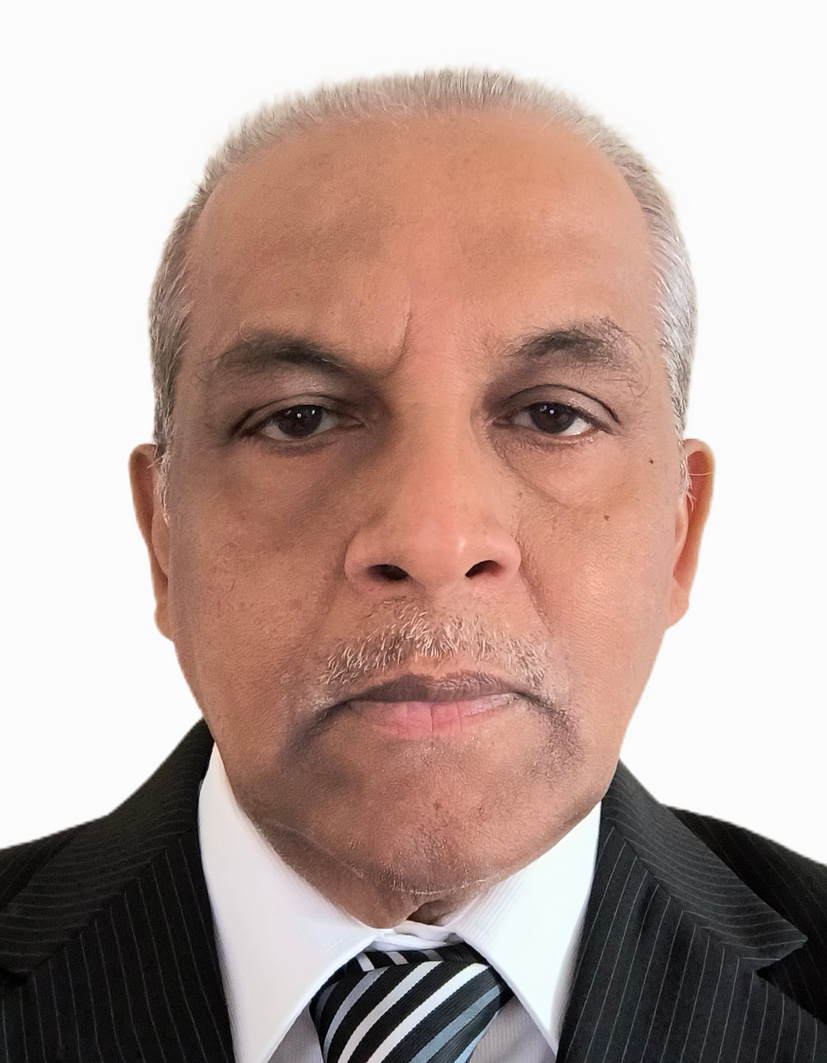}}]%
{Antony Joseph} has been with Philips / NXP semiconductors since 2001 until 2025.
He was a System Architect involved with wireless / wired connectivity and machine learning for edge
applications. Prior to Philips / NXP he was a research engineer at IMEC Belgium,
working on receivers for global positioning systems. He started his engineering career in
radio astronomy building receiver systems for radio telescopes. He hold an MSc degree from
the University of Salford.

\end{IEEEbiography}

\vspace{-0.5cm}
\begin{IEEEbiography}
[{\includegraphics[width=1in,height=1.25in,clip,keepaspectratio]{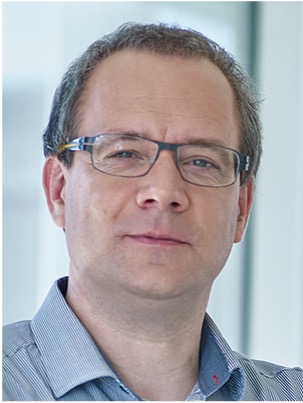}}]%
{Wim Dehaene} obtained his master’s degree and phd from KU Leuven, Belgium, in 1991 and 1996 respectively. Presently, he is a full professor and head of the MICAS division of KU Leuven. His research domain is circuit level design of digital circuits. The current focus is on ultra low power signal processing and memories in advanced CMOS technologies. In memory compute is an integral part of this. Part of Wim Dehaene’s research is performed in cooperation with IMEC, Belgium, where he is also a part time principal scientist. He is also leading the circuit design aspects in several biomedical IC design projects.
Wim Dehaene is a senior member of the IEEE. He is a member of the ESSERC steering committee. He was the general chair of ESSERC 2024. He has also served for several years on the ISSCC program committee.

\end{IEEEbiography}

\vspace{-0.5cm}
\begin{IEEEbiography}
[{\includegraphics[width=1in,height=1.25in,clip,keepaspectratio]{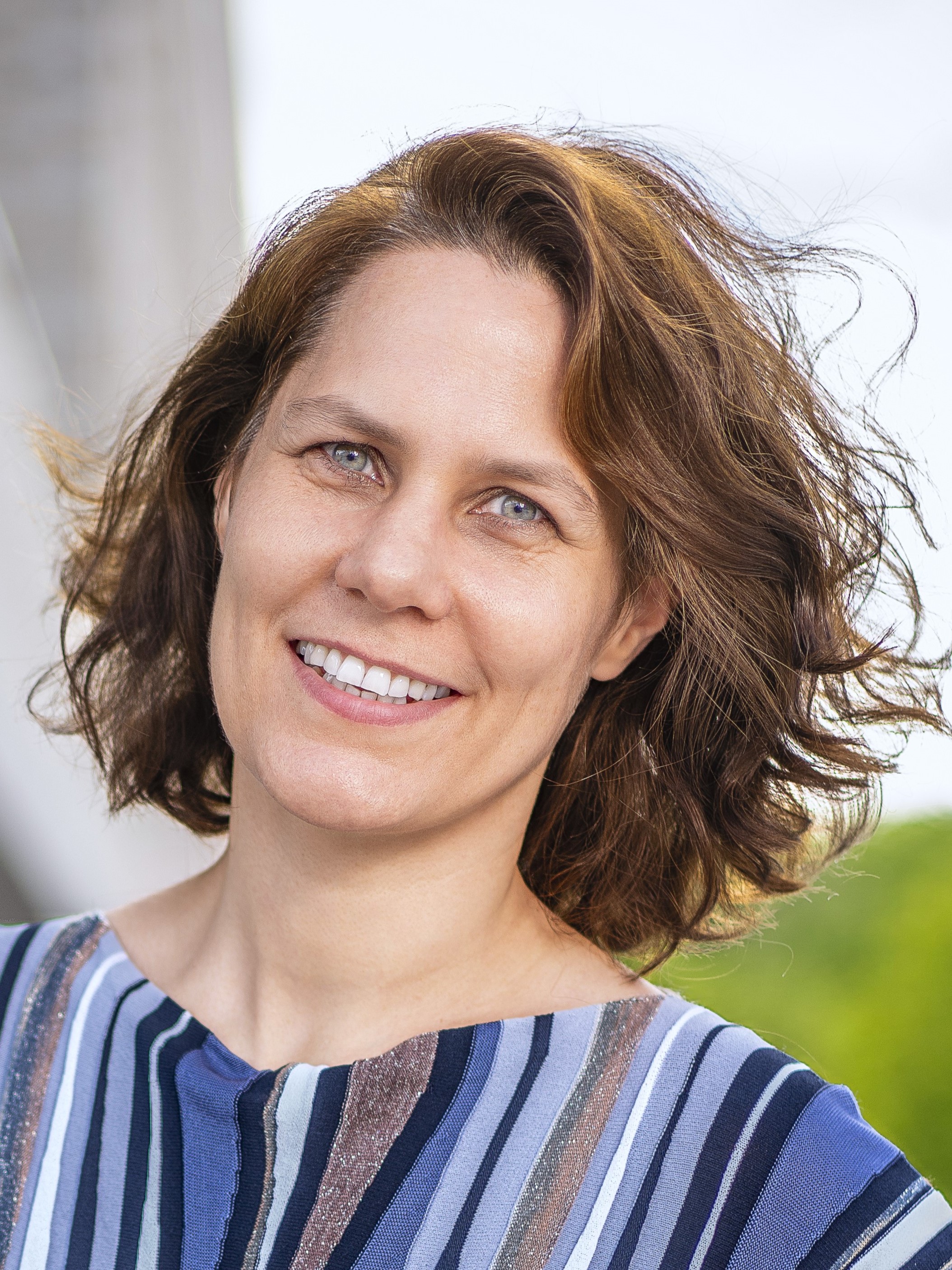}}]%
{Marian Verhelst} is a professor at the MICAS lab of KU Leuven and a research director at imec. Her research focuses on embedded machine learning, hardware accelerators, and low-power edge processing. She received a PhD from KU Leuven in 2008, and worked as a research scientist at Intel Labs from 2008 till 2010. Marian is a scientific advisor to multiple startups, a member of the board of ECSA, and served in the board of directors of tinyML. She is a science communication enthusiast as an IEEE SSCS Distinguished Lecturer, as a regular member of the Nerdland science podcast (in Dutch), and as the founding mother of KU Leuven’s InnovationLab high school program. 
\end{IEEEbiography}
{
\bibliographystyle{IEEEtran}
\bibliography{references}
}
\end{document}